\documentclass{aa}

\usepackage{graphicx}
\usepackage{natbib}
\usepackage[english]{babel}
\usepackage{txfonts}
\usepackage{aalongtable,lscape}

\begin{document}

\title{Coronal abundances of X-ray bright pre-main sequence stars in the Taurus
Molecular Cloud}

\author{L. Scelsi\inst{1,2} \and A. Maggio\inst{2} \and G. Micela\inst{2}
\and K. Briggs\inst{3} \and M. G{\" u}del\inst{3}.}

\offprints{L. Scelsi, \email{scelsi@oapa.astropa.unipa.it}}

\institute{Dipartimento di Scienze Fisiche ed Astronomiche, Sezione di
Astronomia, Universit\`a di Palermo, Piazza del Parlamento 1, I-90134
Palermo, Italy
\and
INAF - Osservatorio Astronomico di Palermo, Piazza del Parlamento 1,
I-90134 Palermo, Italy
\and
Paul Scherrer Institut, W{\" u}renlingen and Villigen, CH-5232 Villigen
PSI, Switzerland
}

\date{Received, accepted}

\authorrunning{L. Scelsi et al.}
\titlerunning{Coronal abundances in Taurus-Auriga X-ray bright stars}

\abstract
{}
{We studied the thermal properties and chemical composition of the X-ray
emitting plasma of a sample of bright members of the Taurus Molecular Cloud to
investigate possible differences among classical and weak-lined T~Tauri stars
and possible dependences of the abundances on the stellar activity level
and/or on the presence of accretion/circumstellar material.}
{We used medium-resolution X-ray spectra obtained with the sensitive EPIC/PN
camera in order to analyse the possible sample. The PN
spectra of 20 bright ($L_{\rm X} \sim 10^{30}-10^{31}$\,erg\,s$^{-1}$) Taurus
members, with at least $\sim 4500$ counts, were fitted using thermal models of
optically thin plasma with two components and variable abundances of O, Ne, Mg,
Si, S, Ar, Ca, and Fe. Extensive preliminary investigations were employed
to study the performances of the PN detectors regarding abundance
determinations, and finally to check the results of the fittings.}
{We found that the observed X-ray emission of the studied stars can be
attributed to coronal plasma having similar thermal properties and chemical
composition both in the classical and in the weak-lined T~Tauri stars. The
results of the fittings did not show evidence for correlations of the abundance
patterns with activity or accretion/disk presence. The iron abundance of
these active stars is significantly lower than ($\sim 0.2$ of) the solar
photospheric value. An indication of slightly different coronal properties in
stars with different spectral type is found from this study. G-type and early
K-type stars have, on average, slightly higher Fe abundances (Fe $\sim 0.24$
solar) with respect to stars with later spectral type (Fe $\sim 0.15$ solar),
confirming previous findings from high-resolution X-ray spectroscopy; stars of
the former group are also found to have, on average, hotter coronae.}
{}


\maketitle

\section{Introduction}
\label{intro}

Studies on the chemical composition of stellar atmospheres are important
to address several issues in astrophysics, such as the evolution of 
stellar galactic populations, the
enrichment of the interstellar medium, the influence of elemental
abundances on star formation processes and on the structure of stellar
interiors through opacity effects. 
In the framework of coronal physics, solar and stellar observations in
the UV and X-ray bands during the last two decades have suggested abundance
anomalies that, if understood, could provide powerful new
diagnostics for the physics of stellar outer atmospheres. In particular,
previous observations of the Sun \citep{Meyer1985,Feldman1992},
showed different chemical abundances in corona with
respect to the photosphere, with a pattern depending on the First Ionization
Potential (FIP) of the various elements. The present view is that elements
with FIP $\lesssim 10$\,eV (\emph{low-FIP elements} Mg, Si, Ca, Ni, Fe) are
overabundant in the solar corona with respect to the photosphere by a factor
$\sim 4-5$; elements with FIP $\gtrsim 10$\,eV (\emph{high-FIP elements} C, S,
O, Ne, Ar, with C and S lying close to the boundary of this definition) have
coronal abundances similar to the corresponding photospheric values. 
This solar-type \emph{FIP effect} is most clearly
present in long-lived active regions, while it is much less marked (or even
absent) in coronal holes \citep{Widing1992,Sheeley1996}.
In order to compare solar and spatially-integrated stellar observations, 
\citet{Laming1994} investigated full-disk solar EUV spectra, 
and detected the FIP effect also in this case.

Analyses of EUV line emission spectra from the coronae of low- and
intermediate-activity stars ($\alpha$\,Centauri, $\xi$ Bootis, $\epsilon$
Eridani, observed with the EUVE satellite) 
revealed solar-like abundance anomalies, suggesting that the 
FIP effect is not an isolated solar phenomenon
\citep{Laming1996,Drake1997,Laming1999}. New studies, based on
XMM-\emph{Newton} and $Chandra$ high-resolution X-ray spectra, have led to
claims of different FIP-dependent abundance patterns in the coronae of other
stars: in particular, observations of very active stars, both single stars and
active binary systems, suggested a dependence of coronal 
abundances on FIP which is reversed with respect to the Sun and the other
low-activity stars mentioned above, i.e. a depletion of low-FIP elements
with respect to high-FIP elements (\emph{inverse FIP effect};
\citep{Brinkman2001,Audard2003,Robrade2005}).
\citet{Telleschi2005} derived coronal abundances for 6 solar analogs (i.e.
G-type main sequence stars) with different ages and rotation periods, and hence
activity levels: this work suggested a change from an inverse FIP effect to a
solar-type FIP effect with increasing age, and hence with decreasing activity. 

On the contrary,
other authors found evidence for neither a FIP nor an inverse FIP
effect in the coronae of active late-type stars
\citep{Sanz2003,Argy2004,Scelsi2005}, but instead a minimum in the abundance
pattern occurring at FIP $\sim 10$\,eV, with increasing abundances at both lower
and higher FIP values. In any case, a systematic result emerging 
from stellar abundance studies
is that the coronae of high-activity stars appear significantly metal-poor
relative to the solar photosphere, with typical iron abundances $\sim 0.1-0.3$
times the \citet{AndersGrevesse1989} solar photospheric value.

A major problem encountered in coronal abundance studies is related to
data analysis issues. The requirements for an accurate analysis are good quality
X-ray spectra (i.e. high $S/N$ ratios), a careful selection of emission lines 
(not affected by blends and independent from the plasma density),
accurate line emissivity functions describing the physics of atomic
transitions, and a precise knowledge of the instrumental response.
All these difficulties introduce errors that are 
often not accounted for in the reported (just statistical) measurement
uncertainties, which therefore are likely underestimated in many studies. 

While the coronal metal deficiency of active stars seems to be a rather solid
result, the question whether coronal abundances are different from the
photospheric values, with a pattern possibly dependent on the FIP, is still
debated and deserves further investigation, especially because
photospheric abundances of active, rapid rotators are mostly unavailable.
Better observational evidence is definitely required to test theoretical
models proposed to explain the physical mechanism(s) responsible for such
differences \citep[e.g. ][]{Schwadron1999,Laming2004}.

In classical T Tauri stars (CTTSs) material from the circumstellar disk
accretes onto the stellar surface and shock waves at the base of the
accretion funnel are expected to heat the gas to X-ray emitting temperatures of
a few million degrees. From recent analyses of XMM-\emph{Newton} and $Chandra$
spectra of pre-main sequence stars, interesting peculiarities in the abundances
of some CTTSs were also noted: TW~Hydrae, TWA~5, and BP~Tauri
show anomalous Ne/Fe abundance ratios 
\citep[$\sim 5-10$ solar,][]{Stelzer2004,Argy2005,Robrade2006}, significantly
higher than those typically found for other stars 
($A({\rm Ne})/A({\rm Fe}) \sim 3$). Such
peculiarities could be inherited from the original cloud where these stars
formed, but they might also be due to some process of chemical segregation 
during the evolution of circumstellar disks. 
In fact, most of the soft X-ray emission of TW~Hydrae and
part of the emission of BP~Tauri have been attributed to accretion
shocks (while the situation for TWA~5 is more uncertain). Iron and other metals
in the disk could condense in grains, as predicted by some models of cloud
chemistry \citep{Savage1996,Charnley1997}. If we suppose that the gas and dust
phases in the circumstellar disk are separated for some reason, with the
metals remaining at larger distance from the star, while neon and other noble
elements in the gas phase accrete more efficiently onto the stellar surface,
then the gas heated by the shock at X-ray emitting temperatures is expected to
be metal-deficient and enriched in neon. The high Ne/O ratio of TW~Hydrae
may be another sign of such a segregation process occurred in the circumstellar
disk of this star, as argued by \citet{Drake2005b}. Should this scenario be
proven correct for the CTTSs and the gas-dust separation mechanism understood,
the chemical composition of the hot plasma could provide new insight into the
study of circumstellar disks. The relevance for our understanding of planetary
formation could also be significant.

Most recently, \citet{Maggio2007} studied a sample of 146 X-ray bright sources
associated with members of the young Orion Nebula Cluster (ONC), and
found that their emission originates from a plasma with temperatures and
elemental abundances very similar to those of active coronae in older
stars. Moreover, the abundance distributions are compatible with a single
pattern vs. FIP for all stars for all elements with the only possible
exception of the Calcium abundance. Comparison with abundances 
in stellar photospheres and in the gaseous component of the nebula
indicate a significant underabundance at coronal level only for iron (by
a factor 1.5--3),
while there is agreement (and hence no evidence of a FIP-related effect)
for all the other elements with available measurements.

In another recent study based on nine high-resolution RGS spectra from the
XEST survey, \citet{TelleschiXEST2006} found similar abundances in accreting and
non-accreting stars and indication of higher Fe abundances in the G-type stars
of their sample compared to stars with later spectral type. They also found that
CTTSs display a soft excess when compared to WTTSs or zero-age main-sequence
stars, and that high electron densities and Ne/Fe abundance ratios are not
characteristic of all accreting PMS stars.

In this paper we analyze X-ray spectra taken with XMM-\emph{Newton} to derive
the thermal properties and chemical composition of the X-ray emitting plasma of
a sample of 20 young stellar objects (7 CTTSs and 13 WTTSs) belonging to the
Taurus-Auriga star forming region and spanning an age range of $\sim 1-10$\,Myr.
This region has been recently the subject of the
XMM-\emph{Newton} Extended Survey of Taurus 
\citep[XEST, ][]{GuedelXEST2006}, which allowed us to study the X-ray emission from more
than 150 classical and weak-lined T Tauri stars in the Taurus Molecular
Cloud (TMC). The present work
focuses on some of the most active members of the TMC, with
the aim to understand whether accreting and non-accreting pre-main
sequence stars show any systematic difference in the abundance pattern,
and to investigate possible dependences of the coronal abundances
on the activity level.

The paper is organized as follows. Section \ref{obs} summarizes relevant
information about the TMC and the XEST observations.
In Sect. \ref{Cap3_abundances} we describe the strategy employed to select
the sample and we report the spectral fitting results, whose
discussion is presented in Sect. \ref{disc}. Finally, we draw our
conclusions in Sect. \ref{concl}. Appendix \ref{sim} details the
simulations performed for this work.

\section{The TMC and the XEST observations}
\label{obs}

The TMC is a cloud complex at 140\,pc from the Sun, which contains about 400 
known low-mass members; they are spread over a relatively large portion of the 
sky ($\sim 100$ square degrees) and their ages cover a range between $10^5$ and
$3\times 10^7$ years, with most stars being typically 1--10 million years old.
The XEST survey consists of 19 recent observations (fields XEST-02 to XEST-20),
obtained in two separate periods (August-September 2004 and February-March 2005)
and with exposures ranging from 27 to 35\,ks, to which 8 more previous
fields\footnote{The field L1495 around V410 Tau was observed twice; the two
observations are labelled XEST-23 and XEST-24.} from the XMM-\emph{Newton}
archive were added, having longer exposure times in the range 43--131\,ks. In
total, the surveyed area covers only $\sim 5$ square degrees of the Taurus
Molecular Cloud; however, since the observed fields were chosen so as to include
the densest regions of the TMC with the largest concentrations of stars, the
total number of detections associated to Taurus members in the EPIC
images is 159. Including non-resolved companions that might be contributing
to the X-ray emission, the XEST survey likely detected around 210 members,
that is about a half of the entire known stellar population in Taurus. This
sample contains mainly classical and weak-lined T~Tauri stars, but also
protostars, brown dwarfs and a few Herbig Ae/Be stars and other members with
uncertain classification.

The present study is based on X-ray spectra taken with the most sensitive 
instrument onboard XMM-\emph{Newton}, the EPIC PN camera
\citep{Struder2001}. This non-dispersive CCD detector has an effective area of
$\sim 1200$\,cm$^2$ at 1.5\,keV and provides moderate spectral resolution 
$R=E/\Delta E \sim 5-50$ in the range $0.1-10$\,keV. Bright sources in the field
XEST-26 (around SU Aur), which lacks PN data, were excluded from the
analysis to ensure homogeneity of the results. For the XEST observations, the PN
was operated in full window mode and the medium filter was applied in all
observations, with the exception of the field XEST-27 (around $\zeta$ Per) where
the thick filter was used. 

We refer to the paper by \citet{GuedelXEST2006} for more details about all the
observations of the XEST project, which were processed using SAS v6.1, and the
data reduction steps. 

\section{Data analysis}
\label{Cap3_abundances}

\subsection{Preliminary simulations and choice of the sample}
\label{Cap3_abund_sample}

A study of the chemical composition and of possible effects due to activity or
accretion processes requires a large sample of stars and accurate
and precise abundance measurements. Accurate estimates of abundances can be
derived with the analysis of high-resolution spectra, where individual
emission lines of several chemical elements are well resolved. The XEST survey
has provided useful RGS spectra for nine TMC members, which have been analysed
by \citet{TelleschiXEST2006}. 

In order to consider a larger sample of T Tauri stars, we based our study on the
medium resolution spectra taken with the PN camera, which is $\sim 10$ times
more sensitive than the RGS spectrometers and, unlike the latter, can also
provide data for sources located off-axis in the field-of-view. The use of PN data required a preliminary feasibility study to understand the reliability
of the results obtained from their analysis. In fact, the blending of the lines
of many elements (neon, iron, nickel, oxygen) and the method used to
analyze these spectra (i.e. global fitting with a plasma emission model
with few isothermal components)
might lead to systematic uncertainties in the
abundances derived for some of the elements. In particular, we addressed
two issues: (i)
Is the global fitting of PN spectra able to recover the true values of
elemental abundances, considering sources in a variety of physical 
conditions? (ii)
What minimum number of detected source counts are required to obtain
measurements with relatively low abundance uncertainties, say within a factor of
2?

These two issues are clearly not unrelated, since the ability of deriving an
accurate and precise abundance value for a given element depends on {\bf a)}
whether the lines of that element have a good signal, and hence on the element
abundance, on the thermal structure of the plasma and also on the amount
of interstellar absorption,
{\bf b)} the statistics of the entire spectrum, which must allow a good
determination of the source thermal properties. We also point out that facing
these issues requires the most accurate knowledge of the instrumental response
function allowed by the instrument calibration.

To answer the above questions, we performed a large number of simulations
exploring a variety of source conditions. At the same time, this
procedure allowed us to put a rough threshold on the total number of source
counts, below which the fitting parameters begin to be poorly constrained, and
hence to define the sample of stars to be studied. The approach is
analogous to that employed by \citet{Maggio2007} for their abundance study on
a sample of stars in Orion, observed with $Chandra$/ACIS.

In brief, our simulations, whose details are reported in Appendix
\ref{sim}, showed that the abundances of most elements can be rather accurately
derived from fitting of PN spectra and set an indicative threshold of 
$\sim 5000$ total counts in order to have useful PN data for abundance studies.
Uncertainties at 90\% level are usually better than a factor 
$\sim$ 2.5 for spectra with $\sim 5000$ counts, and better than a factor
$\sim$ 2 for spectra with $\sim 10\,000$ counts. The iron abundance is the best
constrained in most simulations, although some caveats must be noted in a few
cases (see Appendix \ref{sim}). Instead, the abundances of Ar and Ca are
generally very poorly constrained, as well as the O abundance in case of high
absorption ($N_{\rm H} > 10^{21}$\,cm$^{-2}$).

Therefore, we selected all the TMC members with more than $5000$ counts in the
PN spectra ($0.3-10$\,keV), integrated over time intervals not affected
by high background levels.
We also added to the sample the two
stars Anon 1 and L1551-51 with $\sim 4500$ counts. The sample defined in this
way is listed in Table \ref{tab:lista_stelle_abb} and consists of 20 TMC
members, 13 of which are weak-lined T~Tauri stars, and the remaining 7 are
accreting PMS stars (1 Herbig Be star and 6 CTTSs). Note that the Herbig Be star
in this sample, V892~Tauri, has a close companion ($0.05''$ separation) with an
estimated mass of $1.5-2$\,M$_{\odot}$ \citep{Smith2005}, which therefore is
likely to be the main source of X-ray emission, rather than the Herbig Be star
itself. It is also likely that such a companion is an accreting star, based on
previous studies \citep{Duchene1999,Koenig2001,Hartigan2003} showing that mixed
binary systems, i.e. systems that include both accreting and non-accreting
stellar components, are very uncommon. Using the evolutionary models by 
\citet{Siess2000} and age equal to that of the primary star ($\sim 3$\,Myr),
the spectral type of this companion is K3 for $M=1.5$\,M$_{\odot}$, K2 for
$M=1.7$\,M$_{\odot}$ and K1 for $M=2$\,M$_{\odot}$. In the following we will
assume a spectral type of K2.
\begin{center}
\begin{table}[t]
\caption{Sample of TMC members selected for the abundance study. ``XEST id'' is
the identification label in the XEST catalog, the first two digits referring to
the XEST field. For each source, both the total exposure time of the relevant
observation and the screened time (GTI) selected for this study are reported; 
``PN cts'' are the source counts during the GTIs. ``W'' and ``C'' stand for WTTS
and CTTS, respectively. TTS types and spectral types are taken from
\citet{GuedelXEST2006} and references therein, except where noted.}
\vspace{0.1cm}
\begin{center}
\begin{tabular}{l@{\hspace{1.5mm}}c@{\hspace{2mm}}c@{\hspace{2mm}}c@{\hspace{2mm}}c@{\hspace{2mm}}c@{\hspace{2mm}}c} \hline\hline
Name         &  XEST id & Total    &  GTI  &    PN   &  TTS & Spec \\
             &          &   Exp.   &       &   cts   & type & type \\ 
             &          &   (ks)   &  (ks) &         &      &      \\ \hline
Anon 1       &  20-005  &   26.8   &  26.8 &   4520  &  W   & M0 \\ 
CoKu Tau 3   &  12-059  &   26.4   &  25.7 &  10454  &  W   & M1 \\ 
DN Tau       &  12-040  &   26.4   &  25.7 &   5169  &  C   & M0 \\ 
IT Tau       &  18-030  &   26.8   &  23.8 &   8605  &  C   & K2 \\ 
L1551-51     &  22-089  &   48.9   &  45.4 &   4506  &  W   & K7 \\ 
V773 Tau     &  20-042  &   26.8   &  26.8 &  29971  &  W   & K2 \\ 
V826 Tau     &  22-100  &   48.9   &  45.4 &  16950  &  W   & K7 \\ 
V830 Tau $^a$&  04-016  &   27.6   &  26.1 &   8765  &  W   & K7 \\ 
DH Tau $^a$  &  15-040  &   27.4   &  23.3 &  14780  &  C   & M1 \\ 
HD 283572    &  21-039  &   41.1   &  33.9 &  79424  &  W   & G5 \\ 
HP Tau/G2    &  08-051  &   33.7   &  31.2 &  15033  &  W   & G0 \\ 
Hubble 4     &  23-056  &   62.6   &  51.9 &  16394  &  W   & K7 \\ 
KPNO-Tau 15  &  08-043  &   33.7   &  31.2 &   6017  &  W   & M2.75 \\ 
RY Tau       &  21-038  &   41.1   &  33.9 &   5685  &  C   & K1 \\ 
V410 Tau     &  24-028  &   36.9   &  27.4 &  26740  &  W   & K4 \\ 
V819 Tau     &  23-074  &   62.6   &  51.9 &   5524  &  W   & K7 \\ 
V892 Tau     &  23-047  &   62.6   &  51.9 &  10886  &  C $^c$ & K2 ? $^c$ \\  
BP Tau $^b$  &  28-100  &  116.4   &  88.0 &  17770  &  C   & K7 \\ 
T Tau $^a$   &  01-045  &   67.2   &  37.5 &  29960  &  C   & K0 \\ 
RXJ0422.1+1934 & 01-054 &  67.2   &  37.5 &  11296  &  W    & - \\ \hline 
\multicolumn{7}{l}{$^a$ The light curve suggests that the emission is
affected by the} \\
\multicolumn{7}{l}{\hspace{0.2cm} final stage of a flare decay.} \\
\multicolumn{7}{l}{$^b$ A flare occurred during the observation is excluded
from the} \\
\multicolumn{7}{l}{\hspace{0.2cm} GTIs.} \\
\multicolumn{7}{l}{$^c$ Secondary of the binary system.}
\end{tabular}
\end{center}
\label{tab:lista_stelle_abb}
\end{table}
\end{center}

\subsection{Spectral fittings of TMC members}
\label{Cap3_abund_fitresults}

The PN counts of the selected TMC members were extracted from circular
regions around the sources, having typical radii of $\sim 40''-50''$. For each
source, the background events were extracted from an annulus around the source;
contamination from close X-ray sources within the annulus was also avoided (see
\citealt{GuedelXEST2006} for more details about the calculation of the source
and background extraction regions). The standard tasks of SAS were used to build
the instrumental responses for all sources. In Table \ref{tab:lista_stelle_abb}
we report both the total exposure times of the fields where the selected targets
lie and the Good Time Intervals obtained by excluding from the present analysis 
those time intervals affected by high background emission (due to solar events),
in order to increase the $S/N$ ratios of the studied spectra at high energies;
in the case of BP Tau, a $\sim 30$\,ks interval when the source exhibited a
strong flare was also excluded, so as to avoid possible effects of time
variation of the abundances, as suggested by a number of previous studies
\citep{Tsuboi1998,Guedel1999,Favata1999}. 

\begin{figure*}[t]
\begin{center}
\scalebox{0.35}{
\includegraphics{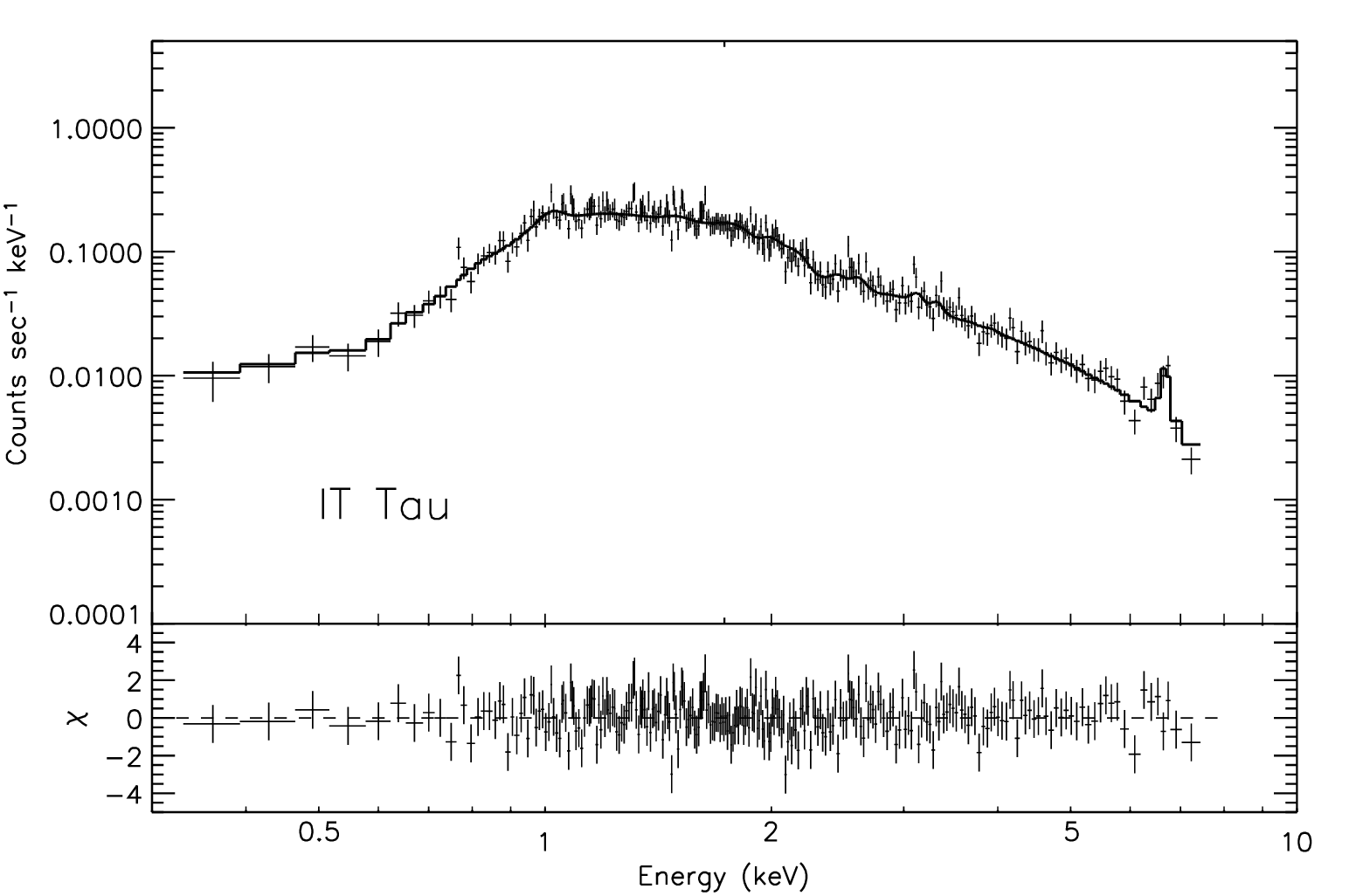}}
\scalebox{0.35}{
\includegraphics{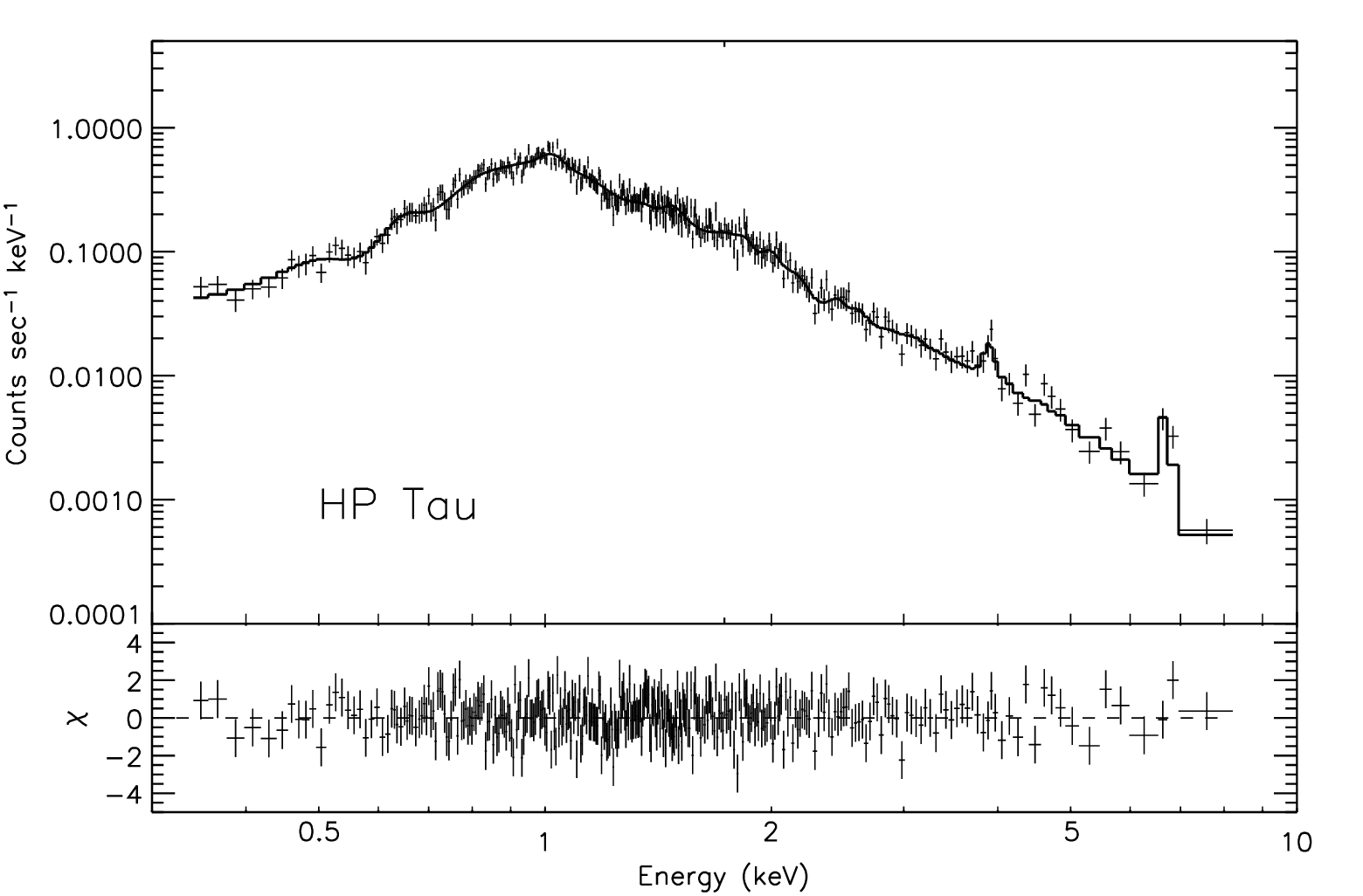}}
\scalebox{0.35}{
\includegraphics{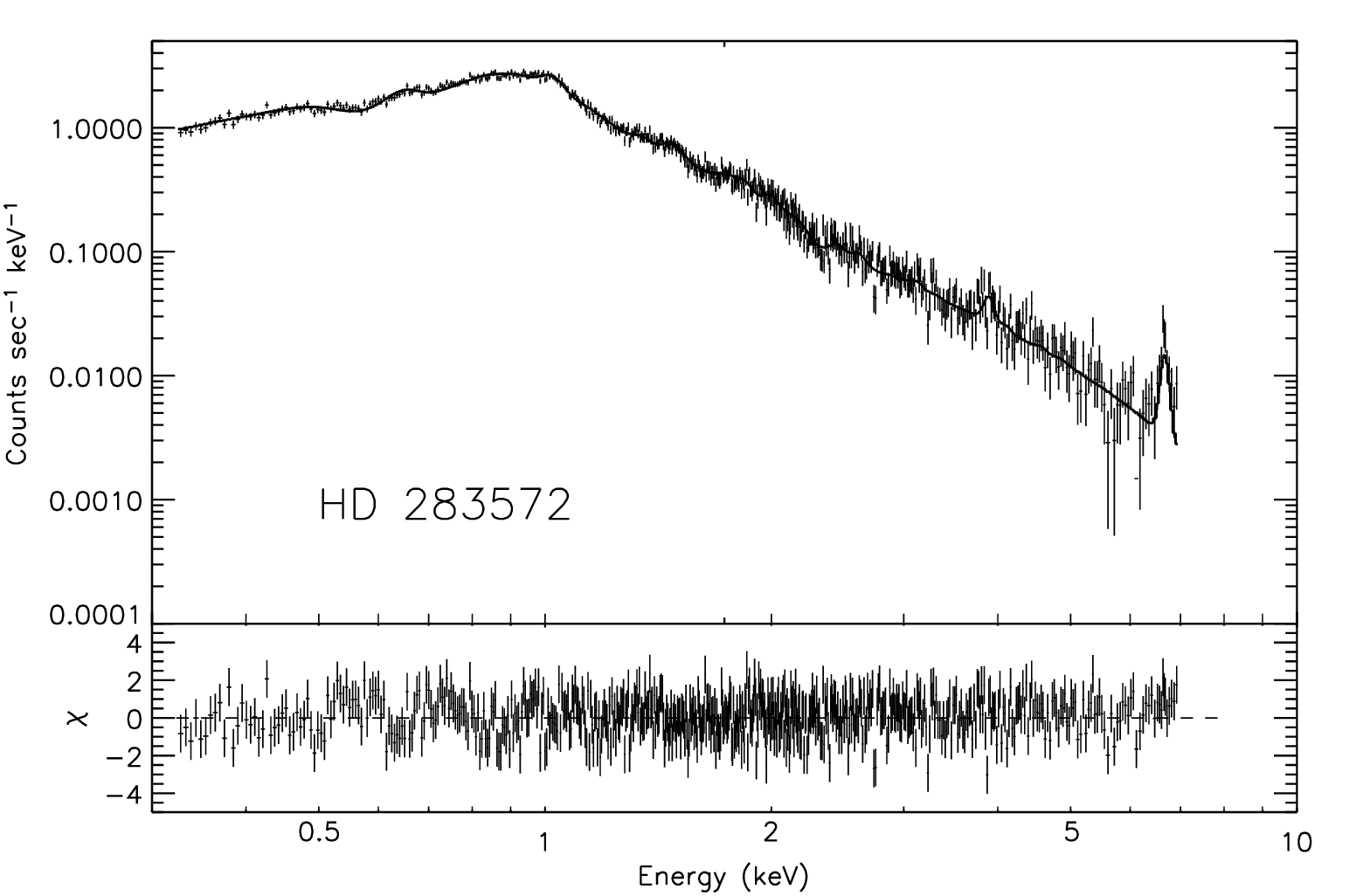}}
\scalebox{0.35}{
\includegraphics{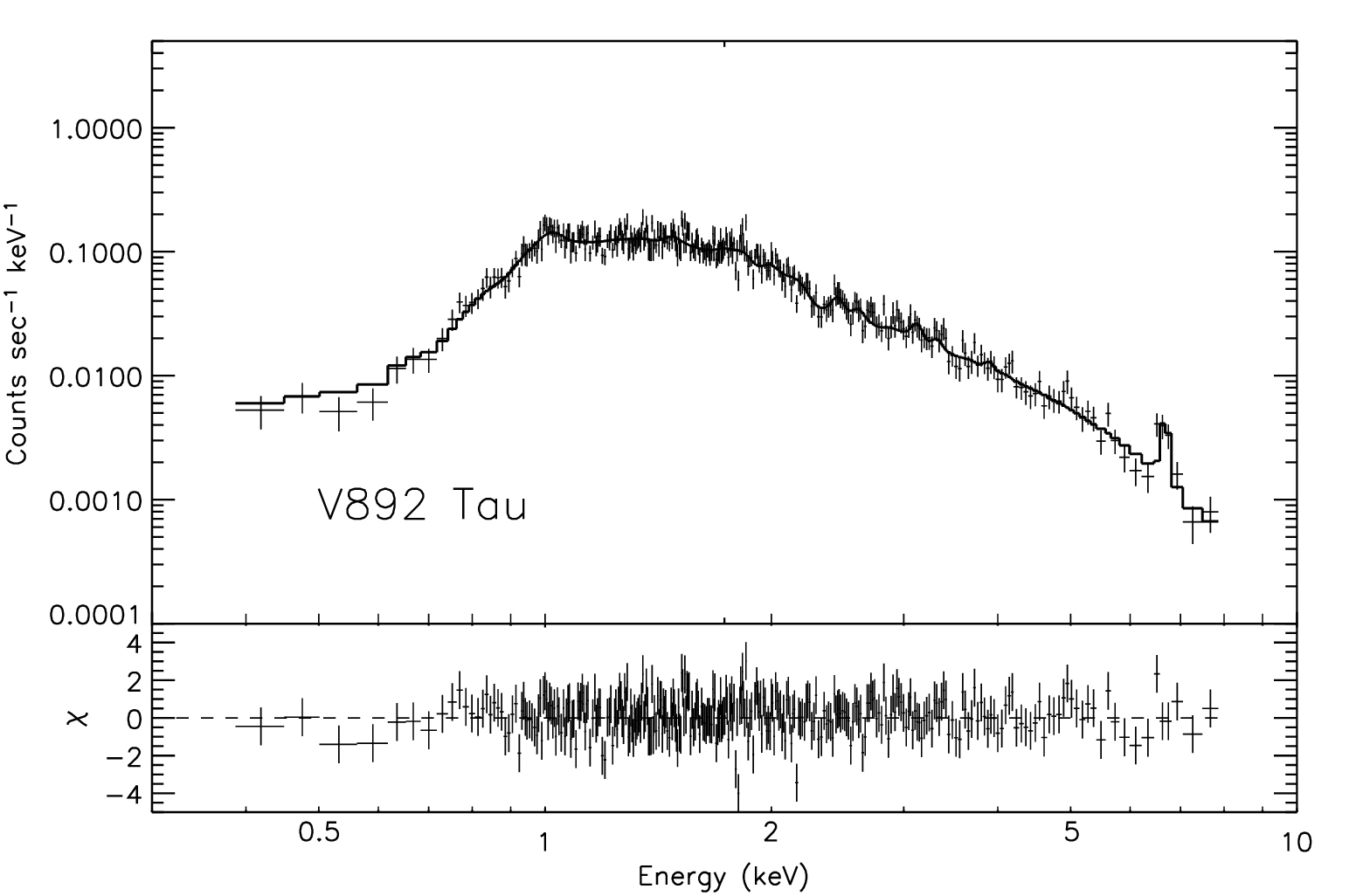}}
\caption{Examples of PN spectra (points with error bars) and best-fit model 
spectra (histogram) of selected TMC members in the present study. The bottom of
each plot shows the residuals in units of $\chi$.}
\label{fig:es_sp_mod}
\end{center}
\end{figure*}

All PN spectra were fitted with the software XSPEC, limiting the fitting of each
source to those energy bins where the background spectrum does not affect
significantly the source emission (usually $0.3-7$\,keV, in very few cases we
excluded the energy bins at $E>5$\,keV). We used a two-component thermal,
optically thin plasma based on the APEC emissivity code, plus interstellar
absorption, which is the same model employed for the simulations. The model has
13 free parameters: the hydrogen column density $N_{\rm H}$, the temperatures
$T_1$ and $T_2$ and the emission measures $EM_1$ and $EM_2$ of the two
components, and the abundances of O, Ne, Fe, Mg, Si, S, Ca and Ar. For each
source, the $N_{\rm H}$ value, the two temperatures and the ratio of the
emission measures of the best-fitting model are reported in Table
\ref{tab:abbond_fit_results}, together with the unabsorbed X-ray luminosity
derived from the model, the average temperature and the surface X-ray flux
derived from $L_{\rm X}$ and the stellar radius (the latter is taken from the
Tables in \citealt{GuedelXEST2006}). Average temperatures are weighted by the
emission measure and are useful to compare the global thermal properties of the
emitting plasma in different stars; they are indicative values and are reported
without errors in the table. The surface fluxes are also reported without
errors, owing to uncertainties in the adopted stellar radii. In Fig.
\ref{fig:es_sp_mod} we plot the observed spectra with their relevant model
spectra, for some example cases.
\begin{center}
\begin{table*}[t]
\caption{Parameters of the best-fit models for the 20 TMC members, with
$1\sigma$ errors. The unabsorbed X-ray luminosity $L_{\rm X}$ and the surface
flux $F_{\rm X}$ are calculated in the $0.3-10$\,keV band. $T_{\rm av}$ is
defined as  $T_{\rm av} = (EM_1 T_1+ EM_2 T_2)\,/\,(EM_1+EM_2)$.}
\vspace{0.1cm}
\begin{center}
\scriptsize
\begin{tabular}{lcccccccc} \hline\hline
Name        &  $N_{\rm H}$  &  $T_1$  &  $T_2$  &  $EM_2/EM_1$ &
$\chi_v2$ (d.o.f.) &   $L_{\rm X}$   &  $T_{\rm av}$   &        $F_{\rm
X}$        \\
            &   ($10^{21}$\,cm$^{-2}$)   &  (MK)   &  (MK)   &            
 &                     & ($10^{30}$\,erg\,s$^{-1}$) &      
(MK)      & ($10^{6}$\,erg\,s$^{-1}$\,cm$^{-2}$)\\ \hline
Anon 1      &  $4.9\pm0.7$  &    $4.8^{+1.1}_{-0.6}$   &   $29^{+22}_{-8}$
   &  $0.12\pm0.06$  & 0.96 (141) & $8.1^{+5}_{-2.2}$ &  7.4  &  10.0 \\
CoKu Tau 3  &  $4.6\pm0.4$  &    $4.3\pm0.7$    &   $18.6\pm1.2$    & 
$1.2\pm0.4$  & 0.88 (284) & $7.4^{+2.0}_{-0.9}$ &  12.2  &  20.6 \\
DN Tau      &  $0.44\pm0.20$ &    $8.7\pm0.8$   &   $24\pm3$    & 
$1.6\pm0.8$  & 1.18 (142) & $1.10^{+0.12}_{-0.05}$ & 18.4  &   3.4 \\
IT Tau      &  $7.0^{+1.2}_{-0.6}$  &    $10^{+3}_{-2}$  &   $35\pm5$    &
 $4.4\pm3$   & 0.92 (256) & $6.6^{+1.8}_{-0.5}$ & 30.7  &  22.8 \\
L1551-51    &  $0.78\pm0.23$ &    $7.4\pm0.7$   &   $21^{+10}_{-4}$    & 
$0.6\pm0.3$  & 1.04 (128) & $1.75^{+0.25}_{-0.09}$ & 12.3  &  14.0 \\
V773 Tau    &  $1.58\pm0.10$  &    $4.8^{+0.5}_{-0.3}$   &   $24.3\pm1.3$ 
  &  $2.0\pm0.3$   & 1.07 (434) & $9.0^{+1.5}_{-0.6}$ & 17.9  &  40.2 \\
V826 Tau    &  $1.03\pm0.11$  &    $4.7^{+0.8}_{-0.4}$   &   $17.9\pm1.6$ 
  &  $0.89\pm0.22$  & 1.16 (304) & $5.1\pm0.3$ &  11.0  &  21.6 \\
V830 Tau    &  $0.79\pm0.20$ &    $6.8^{+1.3}_{-2.0}$   &   $23\pm6$    & 
$0.9\pm0.4$  & 1.17 (237) & $5.7^{+1.0}_{-0.4}$ & 14.5  &  28.7 \\
DH Tau      &  $1.82\pm0.20$  &    $8.9\pm0.5$   &   $25\pm3$    & 
$1.1\pm0.4$   & 1.04 (319) & $7.7^{+0.9}_{-0.4}$ & 17.3  &  37.7 \\
HD 283572   &  $0.55\pm0.05$ &    $8.5\pm0.2$   &   $24.5\pm1.1$    & 
$1.96\pm0.21$   & 1.02 (599) & $12.4\pm0.3$ & 19.1  &  30.7 \\
HP Tau/G2   &  $3.6\pm0.3$  &    $8.7\pm0.5$   &   $22.8\pm1.4$    & 
$1.9\pm0.5$   & 0.94 (329) & $8.1^{+1.1}_{-0.5}$ & 18.0  &  24.1 \\
Hubble 4    &  $2.96\pm0.21$  &    $4.7^{+0.6}_{-0.4}$   &   $19.2\pm1.7$ 
  &  $1.01\pm0.21$   & 1.14 (341) & $6.2^{+0.9}_{-0.4}$ & 12.0  &   9.1 \\
KPNO-Tau 15 &  $3.6\pm0.5$  &    $9.5^{+2.2}_{-1.5}$   &   $33\pm5$    & 
$3.2\pm2.1$   & 0.95 (200) & $2.5\pm0.3$ & 27.0  &  36.7 \\
RY Tau      &  $9.3\pm2.6$  &    $5.3\pm1.6$   &   $35^{+9}_{-6}$    & 
$1.0\pm0.9$  & 0.79 (203) & $8^{+11}_{-3}$ & 19.8  &  10.7 \\
V410 Tau    &  $0.40\pm0.07$ &    $4.7\pm0.4$   &   $20.8\pm1.5$    & 
$1.50\pm0.20$   & 1.04 (364) & $4.93^{+0.26}_{-0.12}$ & 14.4  &  14.9 \\
V819 Tau    &  $1.6\pm0.4$  &    $7.2\pm0.7$   &   $19\pm4$    & 
$0.9\pm0.4$  & 0.95 (175) & $2.3\pm0.6$ & 12.6  &  10.0 \\
V892 Tau    &  $9.2\pm0.8$  &    $10.8^{+1.1}_{-1.6}$  &   $32\pm7$    & 
$0.9\pm0.4$  & 0.94 (325) & $9.0^{+2.5}_{-1.2}$ & 20.8  &  20.7 \\
BP Tau      &  $0.78\pm0.11$ &    $4.7\pm0.4$   &   $22.6\pm2.3$    & 
$1.10\pm0.18$   & 0.99 (338) & $1.45^{+0.12}_{-0.07}$ & 14.1  &   6.0 \\
T Tau       &  $2.53\pm0.16$  &    $9.0\pm0.4$   &   $28.9\pm1.9$    & 
$2.0\pm0.6$   & 1.22 (513) & $7.6^{+0.5}_{-0.3}$ & 22.2  &   9.4 \\
RXJ0422.1+1934 &  $3.4\pm0.3$  &    $4.4\pm0.6$   &   $19.7\pm1.5$    & 
$1.5\pm0.4$  & 0.93 (302) & $3.7^{+0.7}_{-0.3}$ & 13.6  &   -   \\ \hline
\end{tabular}
\end{center}
\label{tab:abbond_fit_results}
\end{table*}
\end{center}
\normalsize

\begin{figure}[t]
\begin{center}
\scalebox{0.5}{
\includegraphics{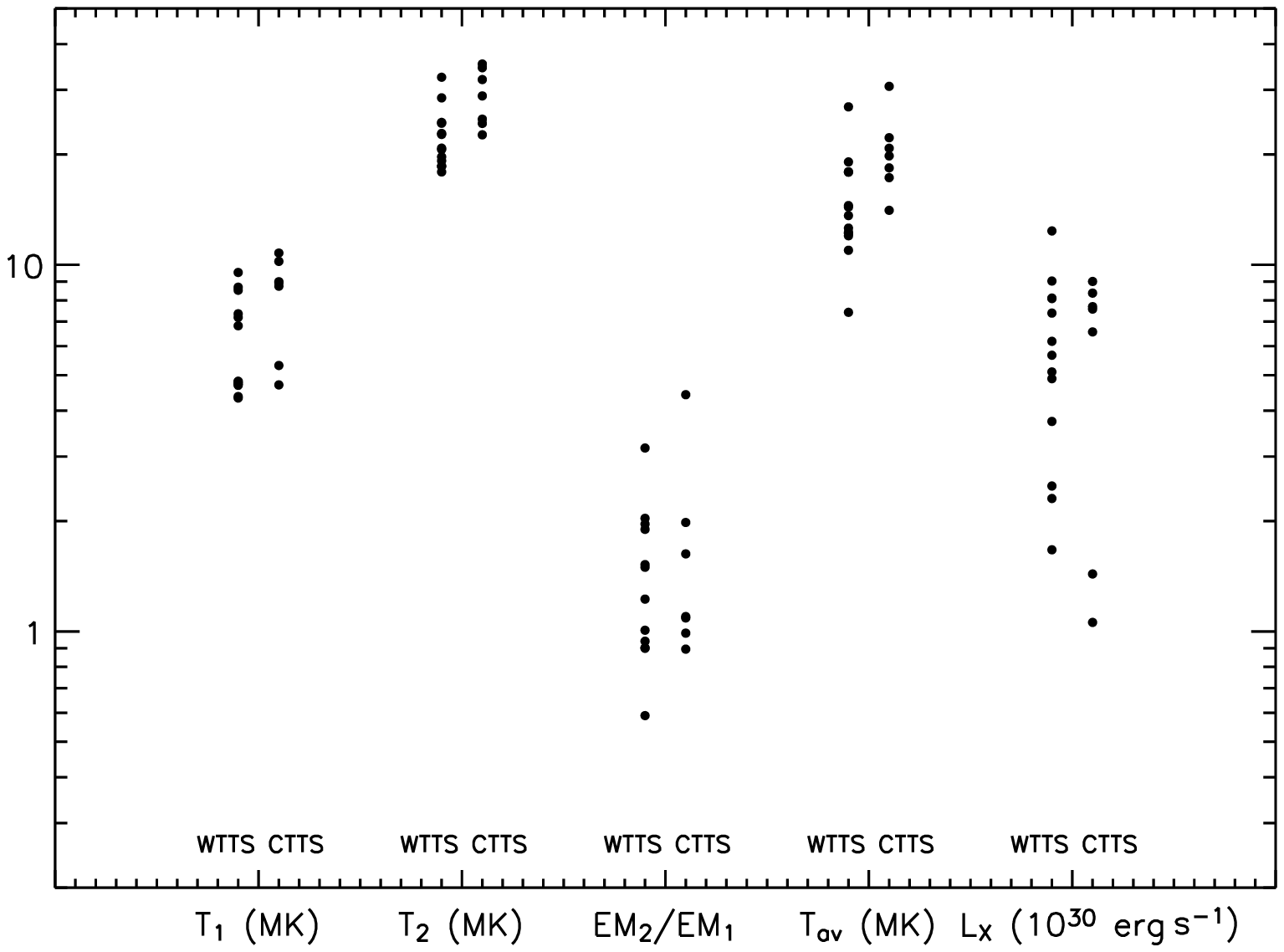}}
\caption{Temperature, emission measure and X-ray luminosity values for the
WTTSs and the CTTSs of the studied sample. The values on the $y$ axis have
different units, for each quantity, as reported in the $x$-axis labels.}
\label{fig:confr_T_EM_Lx_sample}
\end{center}
\end{figure}

Figure \ref{fig:confr_T_EM_Lx_sample} shows the properties of the X-ray
emitting plasma of the studied sample. The distributions of temperatures
and EM ratios are very similar in the subsamples of WTTSs and CTTSs; the
latter seems to have, on the whole, average temperatures slightly greater
with respect to the WTTS (the medians of the distributions of $T_{\rm av}$
are 13.5\,MK and 19.5\,MK, respectively), although a Kolmogorov-Smirnov test
gives 7\% probability that the distributions for WTTSs and CTTSs are not
distinguishable.
\begin{figure}[h]
\begin{center}
\scalebox{0.5}{
\includegraphics{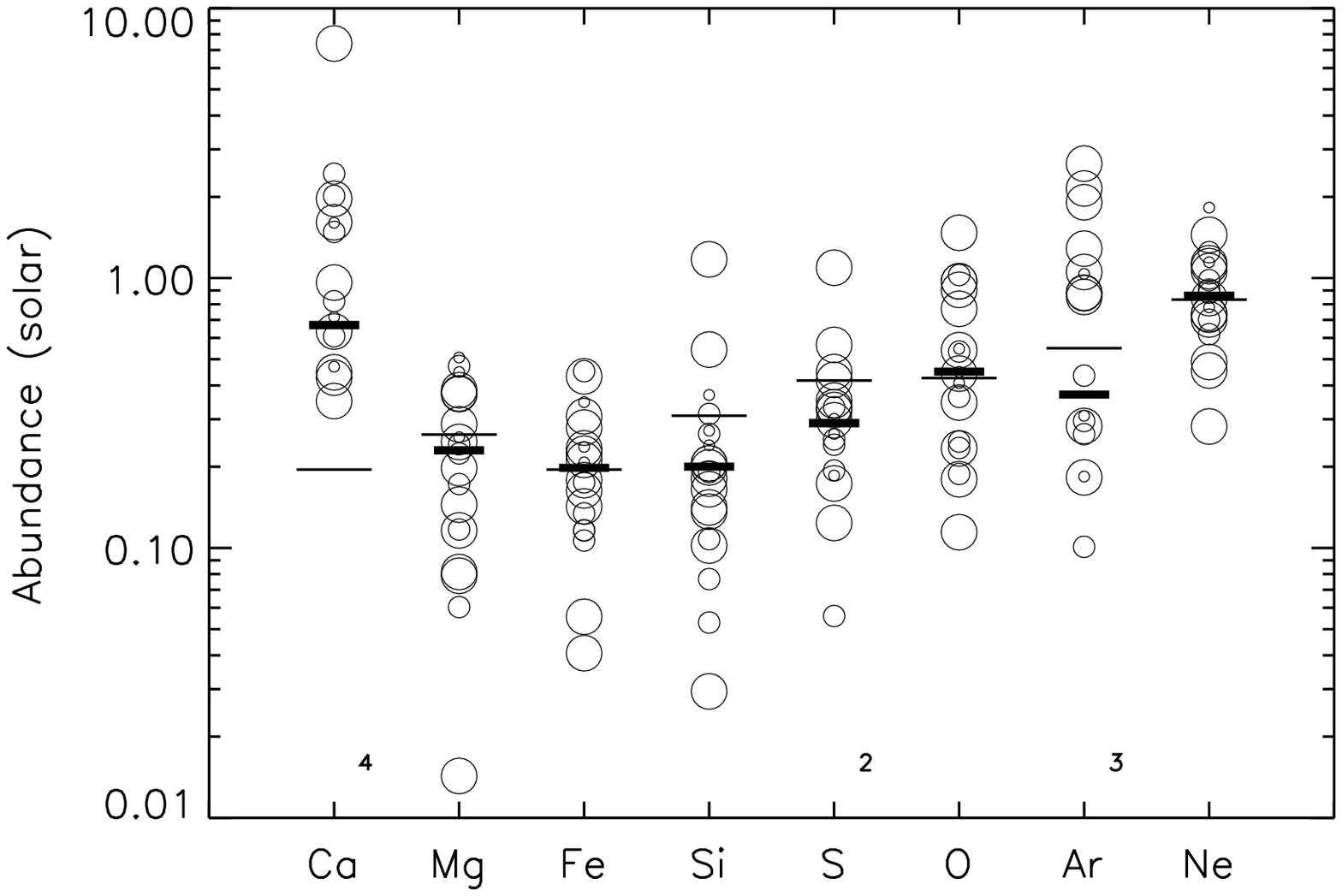}}
\caption{Abundances of the 20 TMC members for each element, ordered for
increasing FIP. The size of the circles gives a rough indication of the
accuracy of the measurements:
large circles for abundances derived from spectra with less than
10\,000 counts, medium circles for $10\,000 <$ counts $< 20\,000$, and
small circles for counts $> 20\,000$. At the bottom of the plot, close to
the label of each element, the number of stars with abundances lying
outside the plot (i.e. lower than 0.01 solar) are indicated. The short thick
segments mark the medians of the abundance distributions, while the long
thin segments mark the values assumed by \citet{GuedelXEST2006} to fit EPIC 
spectra of Taurus members (see Sect. \ref{disc_mediane}).}
\label{fig:abbond_abbond}
\end{center}
\end{figure}

In Fig. \ref{fig:abbond_abbond} we plot the derived abundances (in solar
units of \citealt{AndersGrevesse1989}) of the TMC members, where the elements
are ordered by increasing FIP. The 20 values of abundances are rather widely
distributed. The medians of their distributions show a pattern with a minimum
around Fe and Si, a trend already obtained from the analysis of
high-resolution spectra for several main sequence, pre-main sequence and giant
stars with high X-ray luminosity
\citep{Sanz2003,Argy2004,Scelsi2004,Argy2005,Scelsi2005,Telleschi2005}.
However, as mentioned above, the Ca abundance was generally ill-constrained
in the simulations, which showed that nearly solar values can be
frequently obtained even when the input Ca abundance is significantly subsolar;
hence, if we do not consider calcium, the pattern of the medians
resembles an inverse FIP effect. 

It is important to note, however, that the medians of the abundance
distributions express average properties of the studied sample, while
the abundances of individual stars might be correlated with other stellar
parameters. Yet, comparison of Fig. \ref{fig:abbond_abbond} with the results
of the simulations shows that the spread of the derived abundances for a given
element is roughly of the same magnitude as the one found from the simulations
(in comparing Fig.\ref{fig:abbond_abbond} and Fig.
\ref{fig:esempio_ris_simul} consider that half of our sample stars have less
than $\sim 10\,000$ counts). This implies that \emph{with the
available data} we can hardly distinguish between the abundances of any given
element in different stars of the present sample, and hence it is not meaningful
to search for possible correlations with other parameters. This result also
suggests that the derived sets of abundances of these TMC members could be drawn
from a single set of abundances, i.e. the spread of the values for any given
element is not due to differences among the sources, but rather to statistical
fluctuations. To check this idea, we proceeded as follows: we simulated
10\,000 PN spectra starting from a two-temperature model which reflects the
average observed properties of the studied sample, that is $N_{\rm H}$, $T_1$,
$T_2$, $EM_2/EM_1$ and elemental abundances equal to the medians of the relevant
observed distributions (Table \ref{tab:mediane_2T}). The total number of counts
in the 10\,000 simulated spectra reflects the distribution of counts observed in
the real spectra of the 20 TMC members. Afterwards, we fitted these spectra with
a 2-T model with the same (13) free parameters as before.
\begin{center}
\begin{table*}[t]
\caption{Medians of the distributions of absorption, temperatures, emission
measures and abundances derived from the PN spectra of the studied sample.
Abundances are in solar units, $N_{\rm H}$ in $10^{21}$\,cm$^{-2}$, $T_1$ and
$T_2$ in MK.}
\vspace{0.1cm}
\begin{center}
\begin{tabular}{cccccccccccc} \hline\hline
$N_{\rm H}$ & $T_1$ & $T_2$ & $EM_2/EM_1$ & Ca   &  Mg  &   Fe  &  Si  &  
S  &   O  &  Ar  &  Ne  \\ \hline
    2    &  7  &  23  &     1.2     & 0.67 & 0.23 & 0.195 & 0.20 & 0.29 &
0.45 & 0.37 & 0.86 \\ \hline
\end{tabular}
\end{center}
\label{tab:mediane_2T}
\end{table*}
\end{center}

\begin{figure}[t]
\begin{center}
\scalebox{0.5}{
\includegraphics{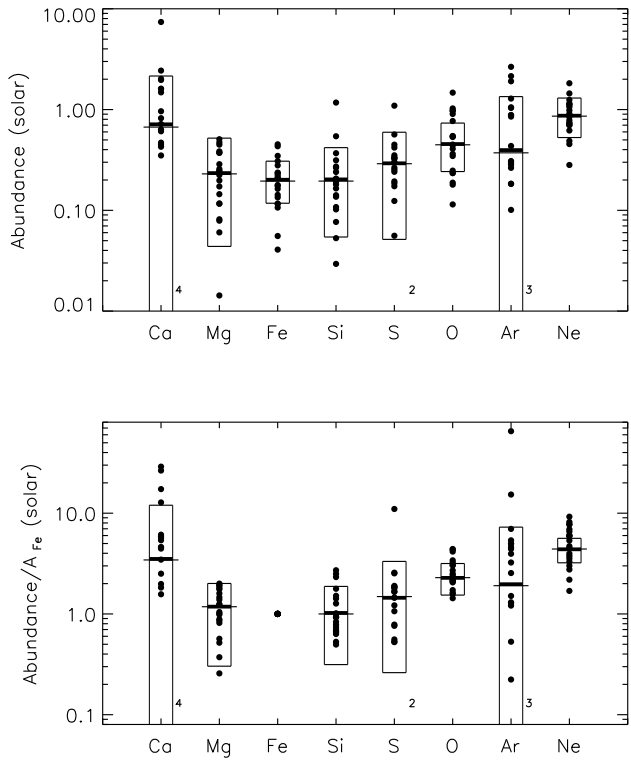}}
\caption{Distributions of abundances (\emph{upper panel}) and abundances
relative to iron (\emph{lower panel}) found for the 20 TMC members, compared
with the results of simulations (see text). The boxes contain the central 90\%
of the abundance values obtained from the simulations, the median being
indicated by the thick black segment within each box. The thin, longer segments
mark the medians of the abundance distributions derived from real data; they are
practically coincident with the median values of the simulations. Numbers at the
bottom of the plot are as in Fig. \ref{fig:abbond_abbond}.}
\label{fig:abbond_test_finale}
\end{center}
\end{figure}

The result of this test is reported in Fig. \ref{fig:abbond_test_finale}, which
shows the distributions of the abundances (in the upper panel) and the 
abundances normalized to iron (in the lower panel) obtained from the real data,
compared with the central 90\% ranges of values obtained from the simulations,
for each element. In both panels, the spread of the values derived from real
data is comparable to that indicated by the simulations, for all elements.
Exception is the Ne/Fe abundance ratio, whose distribution is significantly
larger in the case of the real data. This is better quantified by a K-S test
which shows that for the case of the Ne/Fe ratio the two distributions are
different with a significance of more than $3\sigma$. The range of O/Fe
ratios also appears to exceed the 90\% bounds of the range of the simulations, 
but the difference between the two distributions is not statistically
significant (K-S probability of undistinguishable distributions equal to 7\%).
The result about the Ne/Fe ratio is intriguing and deserves to be further
investigated, because neon and iron are chemical elements having different
properties related to their tendency of forming dust grains, with the volatile
noble gas neon remaining in the gaseous phase while iron more easily condensing
in grains. Moreover, Ne is a high-FIP element, while Fe is a low-FIP element. In
the next Sect. \ref{spread_NeFe} we discuss the Ne/Fe ratio for the stars of the
present sample together with some possibilities that could explain the above
result.

\begin{figure}[!h]
\begin{center}
\scalebox{0.5}{
\includegraphics{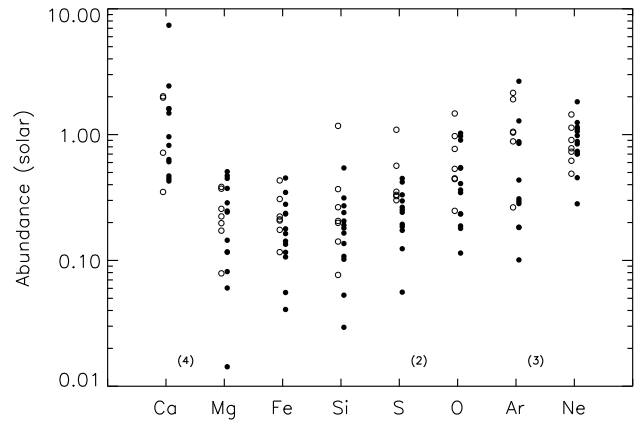}}
\caption{Scatter plot of abundances obtained for the classical (open circles)
and weak-lined (filled circles) T~Tauri stars in the studied sample. Numbers at
the bottom of the plot are as in Fig. \ref{fig:abbond_abbond}.}
\label{fig:abbond_CTTS_WTTS}
\end{center}
\end{figure}

In Fig. \ref{fig:abbond_CTTS_WTTS} we separate the abundance distributions of
the classical and the weak-lined T~Tauri stars: the plot shows that there is no
significant difference between the distributions for the two classes of sources,
indicating that both the thermal properties (Fig. \ref{fig:confr_T_EM_Lx_sample}) and chemical composition of the X-ray-emitting
plasma are similar.

\section{Discussion}
\label{disc}

\subsection{Medians of the abundance distributions}
\label{disc_mediane}

A robust result of this analysis is the low iron abundances of the
X-ray bright Taurus members with respect to the solar photospheric value (Fe 
$\simeq 0.2$ solar), since the abundance of this element is generally
well-determined and with relatively small uncertainties (as indicated by the
simulations). This analysis thus confirms the results of other previous studies
on high-luminosity stars (pre-main sequence stars and RS~CVn binary
systems) which indicate a low coronal metallicity, with respect to the solar
case, in such active stars
\citep[e.g. ][]{Maggio2007,Drake2001,Huenemoerder2001,Audard2003}.

In Fig. \ref{fig:abbond_abbond} we also report the fixed values of 
abundances used by \citet{GuedelXEST2006} for fitting the EPIC spectra of more
than a hundred Taurus members, with the aim of deriving absorption column
densities, characteristic plasma temperatures and X-ray luminosities. 
The values adopted by these authors (see also Appendix \ref{sim}) 
were chosen as representative of the coronal
abundances in active stars. Our detailed analysis confirms the validity
of this assumption, because the
medians of the abundance distributions for all elements are very similar to
the previously adopted values, with the possible exception of calcium.

In Sect. \ref{Cap3_abund_fitresults} we showed that the medians of the 
abundance distributions follow a pattern vs. FIP similar to that derived from
the analysis of high-resolution X-ray spectra for a number of active stars
\citep{Sanz2003,Argy2004,Scelsi2004,Argy2005,Scelsi2005,Telleschi2005};
given the large uncertainties of the Ca abundances, our analysis might
also suggest a systematic pattern of increasing abundances for increasing FIP.
However, such a pattern is obtained by adopting solar photospheric abundances as
reference, while any FIP-related effect should be studied by comparison with the
stellar photospheric abundances; unfortunately, they are not available for any
of the stars in our sample. Measuring abundances in stellar photospheres
is a difficult task, due to the limited quality of stellar optical spectra, to
uncertainties in stellar parameters (gravity, effective temperature,
macro-turbulence) and atmospheric models (LTE approximations), to line blending
(especially for cool M-type stars) and rotational broadening effects 
which is typically the case of active young stars; furthermore, the effects
of possible stellar spots and of the chromospheric activity on abundance
determinations are not well understood, and binarity requires the non-trivial
procedure of untangling composite spectra. An assumption usually adopted is that
stars in the solar neighbourhood have photospheric abundances similar to the
solar ones, because they belong to the same galactic population; moreover, we
expect that stars younger than the Sun have on average higher metallicities
because they have formed from material enriched in high-Z elements by the
remnants of an older stellar population. However, extensive surveys show that
there is a large scatter in metallicity among stars of similar age
\citep{Nordstrom2004}. \citet{Sanz2004} studied the coronal abundances of four
stars with accurately known photospheric abundances and showed indeed that the
pattern of coronal-to-photospheric abundance ratios vs. FIP can be different if
stellar or solar photospheric abundances are used. The recent study by
\citet{Maggio2007} on the X-ray brightest stars in Orion also indicates that a
clear trend vs. FIP is no longer observed when the photospheric abundances of
the Orion stars or the composition of the gaseous phase in the nebula are
considered  instead of the solar ones. On the other hand, \citet{Sanz2003}
derived coronal abundances that are not photospheric for AB~Dor, which has
well measured photospheric abundances, and \citet{Telleschi2005} found the FIP
and inverse FIP patterns in six solar analogs for which several photospheric
abundances are rather clearly known. It is important to dedicate more
observing time to collecting optical spectra in order to measure stellar
photosperic abundances and hence to address in a proper manner the issue of any
FIP-related difference between the photospheric and coronal composition.

\subsection{Ne/Fe abundance ratio}
\label{spread_NeFe}

We now return to the spread of abundances found for the sample stars, for
each element. The only quantity showing a spread of the values larger than
expected from the simulations is the Ne/Fe ratio (Fig.
\ref{fig:abbond_test_finale}), which varies between $\sim 2$ and $\sim 9$, with
a median around Ne/Fe$\sim 5$. The ratio between the abundance of a given
element and that of iron is generally better constrained than absolute abundance
values, because systematic errors in the elemental and iron abundances, due e.g.
to a wrong estimate of the continum emission, usually go in the same direction
and can compensate when calculating their ratio. The result we obtained deserves
to be investigated for several reasons.

A plausible assumption is that the photospheric abundances of iron and neon are
about the same in all the stars studied here, because they belong to the same
region and hence they were born from the same cloud material; then, the Ne/Fe
ratio in their coronal plasma could vary depending on the stellar activity level
if FIP-related effects are really present in stellar coronae, as suggested by
previous studies \citep{Brinkman2001,Audard2003}. 
The X-ray luminosities of the sample stars are in the range
$10^{30}-10^{31}$\,erg\,s$^{-1}$, therefore the study is biased towards active
TMC members. However, \citet{Audard2003} found that a transition from an inverse
FIP effect to a flat pattern with decreasing coronal activity is visible in a
sample of RS~CVn binary systems spanning about the same X-ray luminosity range
as that of the present study.
In particular, they report an anticorrelation between Fe/O vs. the average
coronal temperature $T_{\rm av}$ of five active RS~CVn-like systems as well as a
constant value of the Ne/O ratio, which translates into a correlation between
Ne/Fe and $T_{\rm av}$. When adding a sample of solar analogs studied by
\citet{Guedel2002}, reaching lower activity levels, such a correlation becomes
much more evident and a trend from inverse to normal FIP effect is seen.

\begin{figure}[t]
\begin{center}
\scalebox{0.5}{
\includegraphics{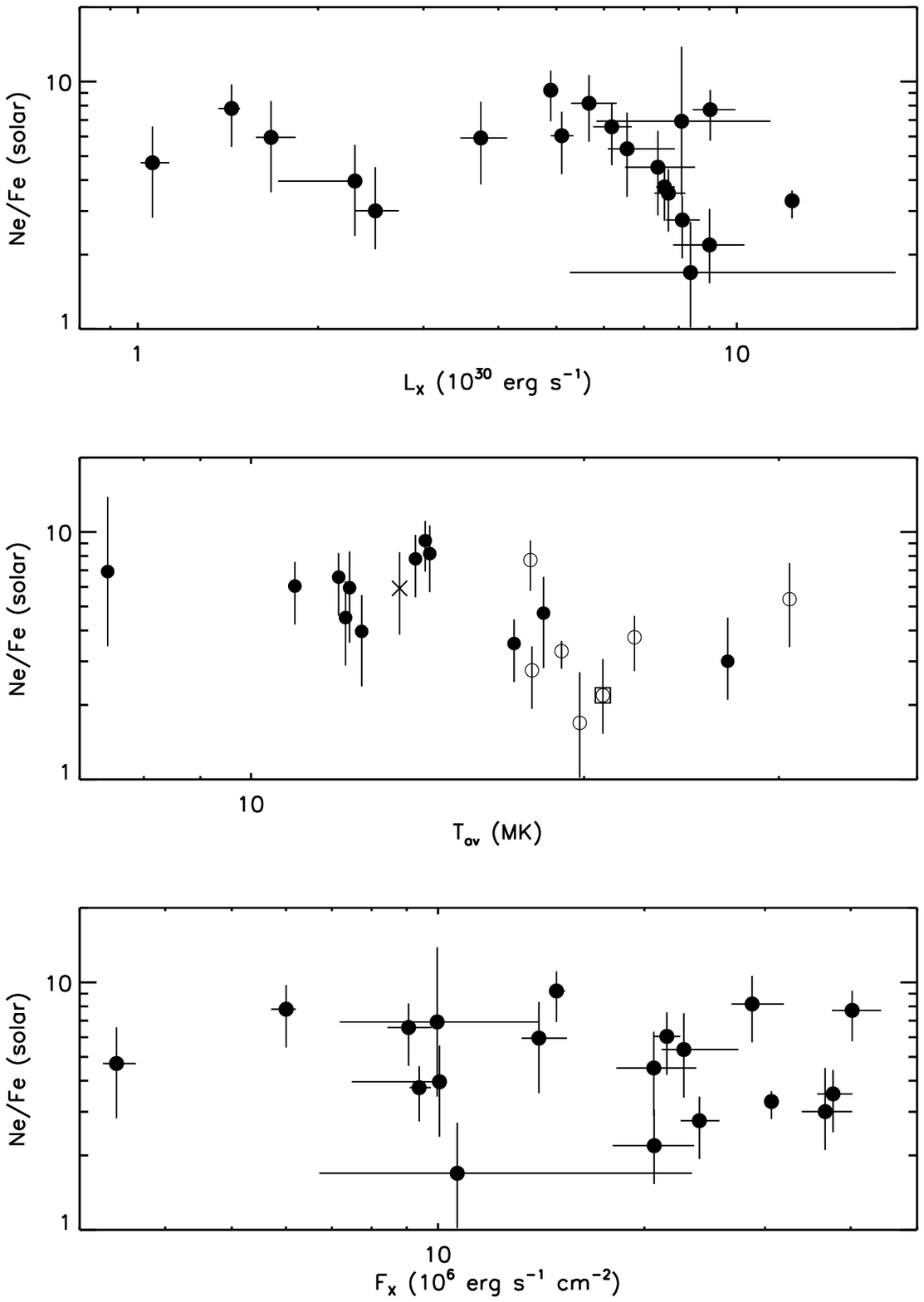}}
\caption{Ne/Fe ratios of all sample stars plotted as a function of three
activity indicators: X-ray luminosity (\emph{upper panel}), average
temperature (\emph{middle panel}) and surface X-ray flux (\emph{lower
panel}). In the plot of Ne/Fe vs. $T_{\rm av}$, different spectral types
are shown (see text): stars with spectral type later than K2 are plotted as
filled circles, while open circles refer to spectral type earlier than K2 (an
open square superimposed to the circle marks the Herbig Be star V892 Tau); the
cross represents the WTTS RXJ0422.1+1934, whose spectral type is not
available.}
\label{fig:NeFe_activity}
\end{center}
\end{figure}

We plot in Fig. \ref{fig:NeFe_activity} the Ne/Fe ratio for our sample of TMC
stars as a function of various activity indicators (X-ray luminosity, 
$L_{\rm X}$, average temperature, $T_{\rm av}$, and surface X-ray flux, 
$F_{\rm X}$), in order to search for possible transitions from a FIP effect to
an inverse FIP effect with increasing activity. However, no significant
correlation of Ne/Fe with any of the considered parameters is found from the
available data, which therefore do not support the existence of different
FIP-related effects in the coronae of the studied stars. 

However, we also note an apparent drop of the Ne/Fe ratios for stars with
average temperature $T_{\rm av} \gtrsim 15$\,MK. Based on the analysis of RGS
spectra of a few stars in the XEST project, \citet{TelleschiXEST2006} also found
a bimodality in the Fe abundance and Ne/Fe ratio for their sample stars,
although this behaviour was observed to be linked to the spectral type. In
particular, the above authors found larger Fe abundances and lower Ne/Fe ratios
in three G-type stars with respect to six K- and M-type stars. To check
their result, we divided our sample into two groups, i.e. stars with spectral
type earlier and later than K2 (7 sources belong to the first group, while 12
stars have spectral type later than K2, the spectral type of the star
RXJ~0422.1+1934 being not available), and we observe in Fig.
\ref{fig:NeFe_activity} that most of the stars cooler than 15\,MK have indeed
spectral type later than K2. The K-S test gives 0.15\% probability that the
distributions of $T_{\rm av}$ for the two groups are drawn from the same
parent distribution. We noted that this difference derives from earlier-type
stars having typically higher $T_2$ and $EM_2/EM_1$ values.

\begin{figure*}[t]
\begin{center}
\scalebox{0.8}{
\includegraphics{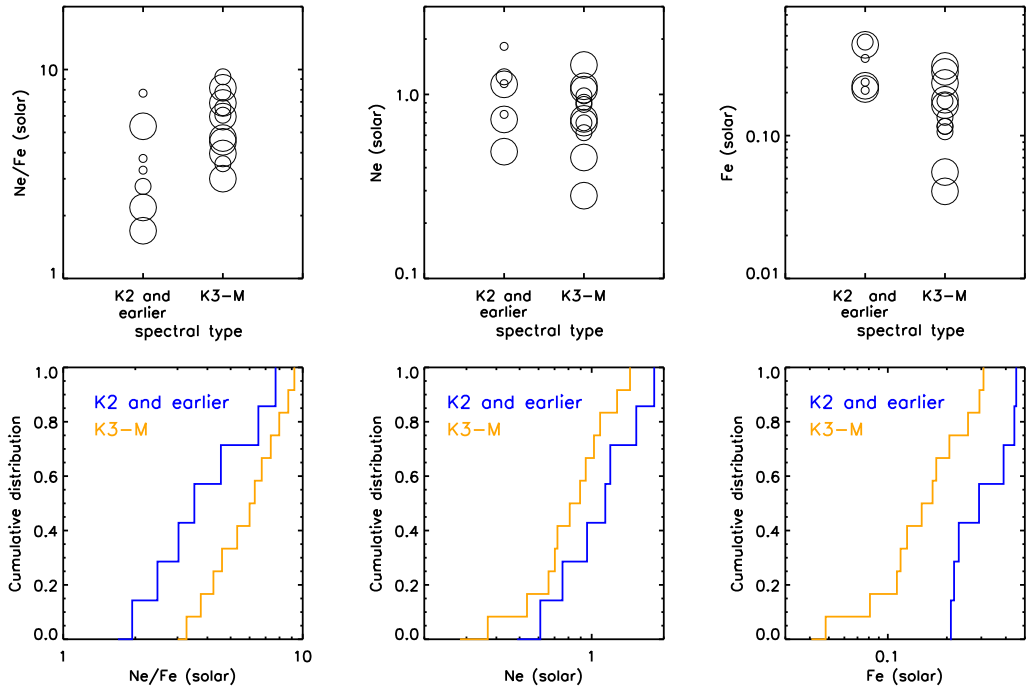}}
\caption{Ne and Fe abundance distributions and Ne/Fe ratios for stars with
spectral type earlier and later than K2 (\emph{three upper panels}); the circle
size decreases with increasing number of total counts in the PN spectrum. In the
\emph{three lower panels} the relevant cumulative distributions are shown.}
\label{fig:NeFe_spectype}
\end{center}
\end{figure*}

Figure \ref{fig:NeFe_spectype} shows the cumulative distributions of the iron 
and neon abundances and of the Ne/Fe ratio for the stars of the two groups:
there is a systematic difference in the neon abundance against the spectral
type, and an even more clear evidence for lower Fe abundances in K- and M-type
stars which implies larger Ne/Fe ratios than in  earlier-type stars. The
probability that the Fe distributions for the two groups are not distinguishable
is 0.6\%; for the cases of the Ne/Fe and Ne distributions,
the probabilities are 8\% and 17\%, respectively. \citet{Franciosini2006}
studied the G-type Taurus member SU~Aur ($T_{\rm av}\sim 20$\,MK) and
found  Fe $\sim 0.6$ times the solar value and Ne abundance about equal to the
solar one, with independent analyses of MOS and RGS data. If we consider also
the abundances obtained for SU~Aurigae in this picture, the result for Fe and
for Ne/Fe is statistically more significant; the three K-S test probabilities
mentioned above are 0.4\% (Fe), 4.5\% (Ne/Fe) and 29\% (Ne). In this
calculation we have assumed that the spectral type of the companion of V928~Tau
(the probable source of X-rays) is K2; we have verified that the above
conclusions do not change assuming for this star a spectral type later than K2.

A possible dependence of Fe and Ne/Fe on the spectral type has not been claimed
in the past, before the report by \citet{TelleschiXEST2006}, and high-quality
spectra of a larger sample of stars are needed to confirm it. Since the stars in
our sample with spectral type later than K2 have, on average, cooler coronae,
this result may also reflect a dependence on $T_{\rm av}$ (although this is not
evident in the Telleschi et al. sample). However, for stars with the highest
$T_{\rm av}$ (in the range $10-30$\,MK), such a dependence would be in the
opposite direction with respect to that found by \citet{Audard2003}, who report
increasing Ne/Fe with increasing $T_{\rm av}$ in the range $\sim 4-15$\,MK.

\begin{figure}[t]
\begin{center}
\scalebox{0.5}{
\includegraphics{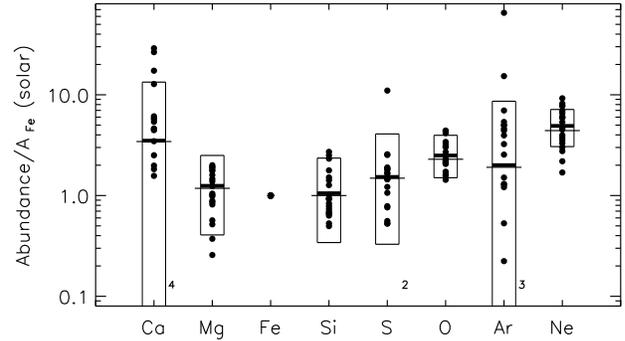}}
\caption{Distributions of abundances relative to iron for the 20 TMC members
compared with the results of simulations taking into account a possible
bimodality of the iron abundance (see text). Symbols and lines are as in Fig.
\ref{fig:abbond_test_finale}.}
\label{fig:abbond_test_finale_G_KM}
\end{center}
\end{figure}

In the light of the above result, we run simulations analogous to those
described in Sect. \ref{Cap3_abund_fitresults}, but taking into account a
possible bimodality of the iron abundance with the spectral type, i.e. we
assumed Fe$=0.24$ and Fe$=0.15$ for stars with spectral type earlier and later
than K2, respectively. Figure \ref{fig:abbond_test_finale_G_KM}, analogous to
the lower panel of Fig. \ref{fig:abbond_test_finale}, shows that the spread of
Ne/Fe in real data can be explained in this way (K-S test probability $= 32$\%).

Another issue of potential importance is that Fe forms grains much more readily
than Ne: as explained in Sect. \ref{intro}, if iron settles in
dust grains in the disk of CTTSs and the dust and gas components are separated
for some reason, X-ray emission from accreting stars may reveal a metal-depleted
plasma. We note that such a result may be expected if
accretion-driven emission dominates over a coronal component, since in this case
the abundances we are measuring are essentially those of the accreted plasma.

\begin{figure}[t]
\begin{center}
\scalebox{0.55}{
\includegraphics{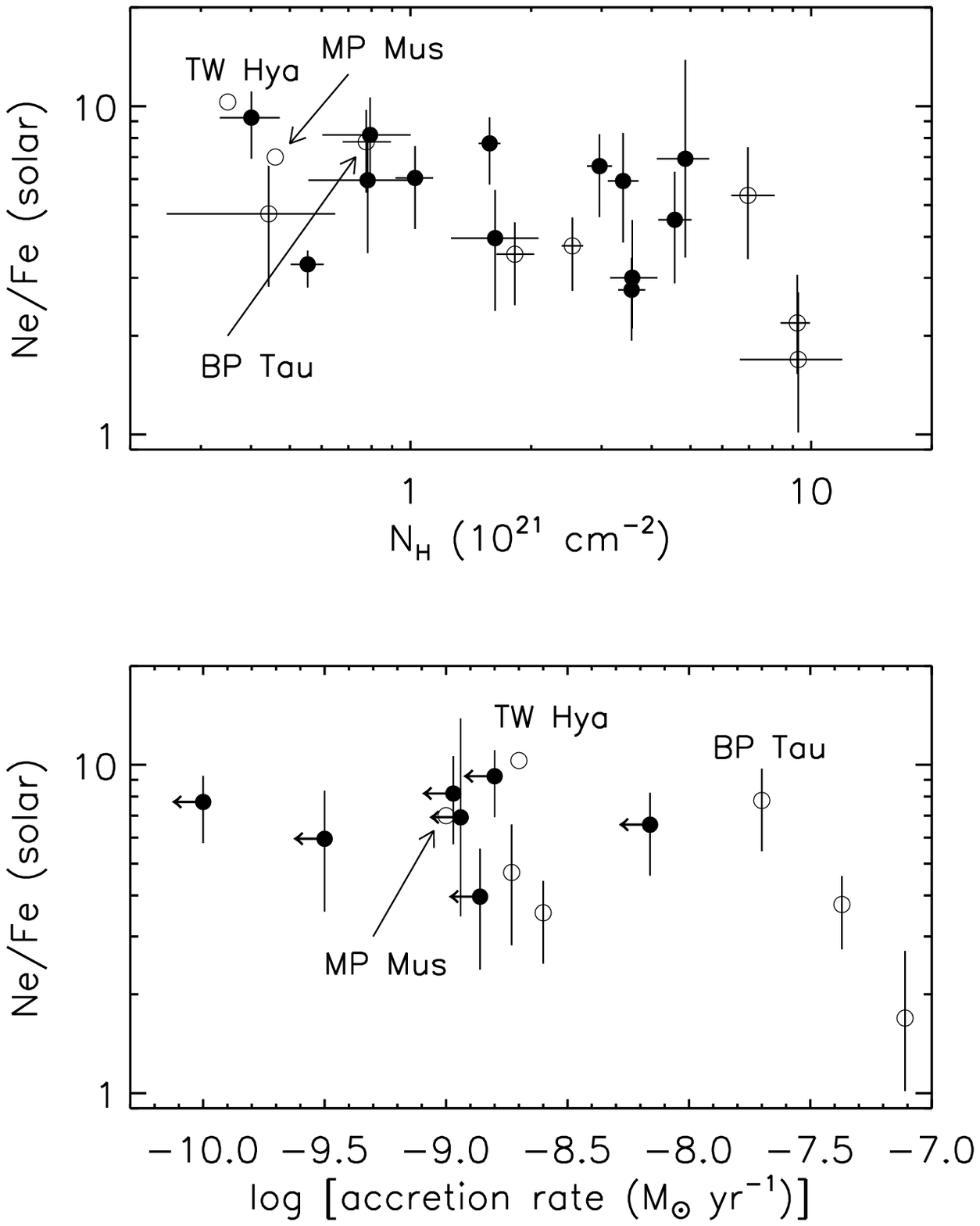}}
\caption{Ne/Fe ratios as a function of the photoelectric absorption
(\emph{upper panel}) and accretion rate (\emph{lower panel}). Errors are
$1\sigma$. Filled and open circles mark WTTSs and CTTSs respectively. In
both plots, the positions of the two CTTS TW Hydrae and MP Muscae are shown as
open circles without error bars. In the lower panel, only stars with measured
accretion rates are shown; all WTTSs have upper limits indicated by left arrows.
Accretion rates for the TMC members, for TW Hydrae and for MP Muscae are
taken from the tables in \citet{GuedelXEST2006}, from \citet{Alencar2002} and
from Sacco G.~G. (private comunication), respectively.}
\label{fig:NeFe_env}
\end{center}
\end{figure}

The X-ray emitting plasmas of the stars in our sample have rather high average
temperature ($\gtrsim 10$\,MK), and similar properties in both the CTTSs and
the WTTSs, indicating that we are probably seeing a dominant contribution from a
coronal emission. Figure \ref{fig:NeFe_env} shows the Ne/Fe ratio of our
sample stars as a function of two stellar parameters linked to the presence of
circumstellar material, i.e. $N_{\rm H}$ and the accretion rate. For
comparison, we report also
the ratios for the two classical T~Tauri stars TW~Hydrae \citep{Robrade2006} and
MP~Muscae \citep{Argy2007}; for the CTTS BP~Tauri, previously studied by
other authors, we note that our Ne/Fe ratio is in very good agreement
with that found by \citet{Robrade2006}. As expected, we do not find evidence of
any relation between Ne/Fe and these two parameters; moreover, as shown in
Fig. \ref{fig:NeFe_env}, high Ne/Fe ratios may be present both in CTTSs and in
WTTSs and are therefore not a peculiar characteristic of accreting stars. This
is consistent with the findings of \citet{TelleschiXEST2006} that high Ne/Fe
ratios are not present in all accreting stars.

\subsection{Ne/O ratio}
\label{NeO_ratio}

\begin{figure}[t]
\begin{center}
\scalebox{0.6}{
\includegraphics{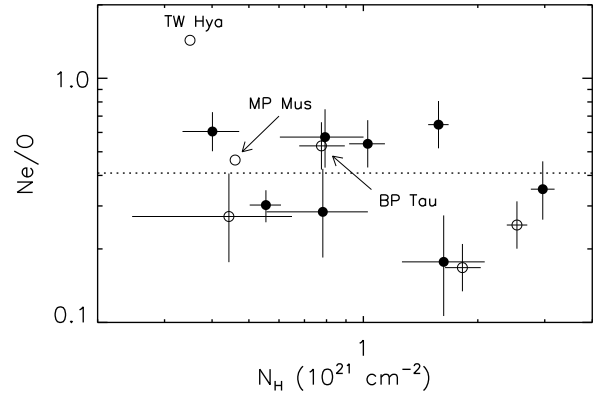}}
\caption{Ne/O ratios as a function of the photoelectric absorption, for
stars with $N_{\rm H} \le 3\times 10^{21}$\,cm$^{-2}$. Errors are $1\sigma$.
Filled and open circles mark WTTSs and CTTSs, respectively. For comparison, the
positions of the two CTTSs TW Hydrae and MP Muscae are also shown (open circles
without error bars). The dotted line marks the average value of 0.41 found
by \citet{Drake2005} for stars of various activity levels.}
\label{fig:NeO_env}
\end{center}
\end{figure}

The Ne/O ratio shares some similarities with the Ne/Fe one, in that oxygen can
easily combine with dust grains; \citet{Drake2005} analyzed high-resolution
$Chandra$ spectra of a variety of stars (dwarfs, giants, multiple systems,
subgiants) and found that the Ne/O ratio is remarkably constant in stellar
coronae (Ne/O $\sim 0.4$, the solar value by \citealt{AndersGrevesse1989} being
$\sim 0.14$). This is true also for the classical T~Tauri stars BP~Tauri and
TWA~5, but not for the CTTS TW~Hydrae, whose higher Ne/O value ($\sim 1$) was
interpreted by \citet{Drake2005b} in terms of depletion of grain-forming
elements in the X-ray emitting gas which is accreting onto the star: such
a depletion is supposed to be more pronounced in the case of the more evolved 
and older
circumstellar disk of TW~Hydrae, where coagulation of grains into larger bodies
orbiting the star could have happened; in the younger CTTS BP~Tauri, where part
of the emission is accretion-driven, the dust particles could be still small and
can release, after heating, the elements locked into the grains. However, very
recently \citet{Argy2007} obtained Ne/O $\sim 0.5$ for the CTTS MP~Muscae,
which is even older than TW~Hydrae, but still shows evidence for a substantial
contribution to the X-ray emission from accretion shocks in the high plasma
density derived from  the analysis of the O {\sc vii} triplet; the average
plasma temperature of this  star is relatively high and
hence the derived abundances are likely affected by the coronal emission. What
is also emerging from these abundance studies is the need for a criterion to
understand which abundances (coronal or accretion-related) we are measuring in
these stars, perhaps also with the help of modeling. 
For the stars we studied here,
the Ne/O ratio is affected by large uncertainties in those sources with high
absorption, because of the difficulty of constraining the O abundance in this
case. If we consider the 12 stars with $N_{\rm H}\le 3\times10^{21}$\,cm$^{-2}$,
we find that the Ne/O ratio is in the range $\sim 0.2-0.6$, with the values
equally spread around $\sim 0.4$ (Fig. \ref{fig:NeO_env}). In particular, this
is true also for the classical T~Tauri stars DN~Tauri, DH~Tauri, T~Tauri and
BP~Tauri\footnote{The value Ne/O found in this study for BP~Tauri is in
good agreement with that obtained by \citet{Robrade2006}.}, among the 12 we are
considering. They are rather young, with ages in the range $\sim 1-2.7$\,Myr,
and their high average plasma temperatures imply a large coronal contribution,
as already discussed. Therefore, our result does not go against the scenario
mentioned above.

\section{Summary and conclusions}
\label{concl}

The thermal properties and chemical composition of the X-ray emitting plasma in
a sample of bright Taurus members have been studied in this work with the aim of
investigating possible systematic differences between classical and weak-lined
T~Tauri stars, and of understanding whether the pattern of abundances in such
stars is related to some stellar or environmental characteristics. We used data
from the XEST survey; since there are very few useful RGS spectra of Taurus
members, we used the PN data in order to enlarge as much as
possible the sample to be analyzed. Preliminary extensive simulations showed
that PN spectra with $\sim 5000$ counts or more in the $0.3-10$\,keV band can
provide reliable estimates of elemental abundances and 68\% statistical
errors usually within a factor $\sim2$ for most of the elements; exceptions may
be O, Ca and Ar in some cases, depending on the statistics and source
absorption. The sample of stars considered for this work therefore consists of
20 TMC members: 13 weak-lined T~Tauri stars and 7 accreting stars, 
having at least 4500 total counts in the
spectrum. Their medium-resolution spectra were globally fitted with an absorbed
two-temperature thermal model with variable abundances of O, Ne, Mg, Si, S, Ar,
Ca and Fe.

The temperatures of the two components were found in the ranges $\sim 5-10$\,MK
and $\sim~20-40$\,MK, and the hot-to-cool emission measure ratios range between
$\sim 0.8$ and 4. Average temperatures are in the range $\sim 8-30$\,MK, with
the classical T~Tauri stars being slightly hotter than the WTTSs.

The spread of the 20 abundance values for each element 
(Fig. \ref{fig:abbond_abbond}) is more or less compatible with the errors
obtained from the simulations; \emph{a posteriori} tests, also based on
simulations, have shown that our results are consistent with these TMC stars
sharing a unique set of abundances, with the possible exception of iron.

The main conclusions from the present work are summarized as follows.

Hot coronal plasma is responsible for most of the X-ray emission from both CTTSs and WTTSs studied here. The coronal plasma has similar properties in the two
classes of T Tauri stars, in particular similar characteristic temperatures and
chemical composition. Late-type stars therefore develop hot coronae early in
their life (within $\sim 1$\,Myr), whose properties change little during the
following pre-main sequence evolution, at least up to an age of $\sim 10$\,Myr.
The reason why the coronal emission appears suppressed in some CTTSs, such as
TW~Hydrae, and not in most of the others is to be clarified with future work.

The available data do not allow us to evaluate quantitatively the presence, in
the CTTSs considered here, of a cool X-ray emission excess that might be present
due to shock-heating.

The iron abundance is significantly subsolar ($\sim 0.1-0.5$ times the solar
photospheric value) in all the sample stars. Since the abundance of this element
is the best-constrained one, this result is rather robust and confirms previous
finding for PMS and active stars. 

We found indication of coronae in G-type and early K-type stars hotter than in
stars with later spectral type. We also find strong support for a
bi-modality of the Fe abundance (and the Ne/Fe ratio), depending on the spectral type (or perhaps on the average temperature), with higher Fe values and lower
Ne/Fe ratios for G-type and early K-type stars compared to stars of later
spectral type. The latter result supports the findings from high-resolution
spectroscopy by \citet{TelleschiXEST2006}. A larger sample is needed, however,
to find out if this trend is merely the result of the limited statistics
presently available.

The analysis of the abundance distributions of this sample of bright TMC members
did not reveal any significant difference in the abundances related to the
presence of accretion/circumstellar material. Ne/Fe abundance ratios between 2
and 9 times the solar value are found both for the classical and the weak-lined
T~Tauri stars, confirming the findings by \citet{TelleschiXEST2006} based
on a smaller sample of high-resolution X-ray spectra of T~Tauri stars.
Regarding the Ne/O ratio, we do not observe in the classical T~Tauri stars a
behaviour analogous to that of TW Hydrae, all values being below $\sim 0.6$.

\begin{acknowledgements}

We acknowledge financial support by the International Space Science
Institute (ISSI) in Bern to the XMM-{\it Newton} XEST team. The Palermo
group acknowledges financial contribution from contract ASI-INAF
I/023/05/0. X-ray astronomy research at PSI has been supported by the
Swiss National Science Foundation (grants 20-66875.01 and 20-109255/1).
This research is based on observations obtained with XMM-{\it Newton}, an
ESA science mission with instruments and contributions directly funded by
ESA member states and the USA (NASA).

\end{acknowledgements}

\bibliographystyle{aa}
\bibliography{scelsi2007}

\appendix

\section{Simulations}
\label{sim}

In this section we describe the simulations we run before fitting the real data
in order to assess the performances of the PN regarding abundance
determinations. This part of the investigation can be summarized as follows:

\begin{enumerate}
\item{Fourteen input coronal models were chosen, each consisting of an EM
distribution and a set of elemental abundances, plus interstellar absorption.}
\item{Three different numbers of source counts were considered: $\sim$\,5\,000,
10\,000 and 20\,000 counts in the $0.3-10$\,keV band of the PN spectrum.
This choice is motivated by the fact that most of the brightest TMC
members have total PN counts within the above range; moreover, previous fittings
of Taurus members with less than 5000 counts in the PN, using 2-T models with
a fixed set of abundances \citep{GuedelXEST2006} already provided good
descriptions of the PN spectra, a result which implies no need to treat the
abundances as free parameters.}
\item{For each simulated observation, i.e. for each coronal model at each
number of total counts, 1000 PN spectra were simulated (using the
standard PN response function and assuming no background contamination) which
have a Poissonian distribution of the total number of counts.}
\item{For each simulated observation, the 1000 PN spectra were
fitted with an automatic routine, using a 2-T coronal model with the following
starting values for the fitting parameters: abundances and $N_{\rm H}$ equal to
the values of the input coronal model, $T_1$ and $T_2$ given by the results of
interactive fittings of a few simulated spectra for each of the fourteen input
models.}
\end{enumerate}

Moreover, for the input models 1 and 2 (see Table \ref{tab:modelli_input_simul}
below) the fitting of simulated spectra was repeated for each of the three
values of the total counts, starting from different initial values of 
$N_{\rm H}$ and temperatures, and initial abundances equal to the solar values:
for each of these 6 simulations, 300 spectra were automatically fitted by
performing an extensive exploration of the $N_{\rm H}$, $T_1$ and $T_2$ space so
as to simulate the approach to real data. The results obtained in this way are
practically undistinguishable from those derived with the procedure
described in item \#4 above. For the input models 9 and 10, non-solar
abundances as starting values for the fitting of simulated spectra were also
used. Again, the results in these cases are very similar to those obtained with
the procedure in item \#4.

Note that using the same instrumental response both to simulate and to fit the
spectra implies that this investigation leads to estimate the errors in the
determination of the plasma parameters due to the photon counting statistics
and to the instrument spectral resolution. The method cannot provide
estimates of possible systematic errors due to uncertainties in the instrumental
calibration and in the atomic parameters used to calculate theoretical plasma
emissivities.

The input coronal models were chosen so as to explore characteristic absorption
values and physical conditions of the coronal plasma (in terms of EMD and
abundances), among those commonly found in the literature for active T~Tauri
stars. For the abundances, some extreme cases were also explored (see below).
The emission measure distributions we have considered for these simulations
are defined in the temperature range $10^6\,{\rm K} \le T \le
10^{7.5}\,{\rm K}$ as:
\begin{equation}
EMD \propto\left\{ \begin{array}{ll}
               T^{\alpha}  & \mbox{\hspace{2cm}$10^6\,{\rm K}\le T \le
T_{\rm peak}$} \\
               T^{-\beta}  & \mbox{\hspace{2cm}$T_{\rm peak}\,{\rm K}\le T \le
10^{7.5}\,{\rm K}$}
                         \end{array}
          \right.
\end{equation}

Based on EMD reconstructions with high-resolution spectra of intermediate- and
high-activity stars by
\citet{Laming1999,Argy2004,Scelsi2005,Argy2005,Telleschi2005,GuedelXEST2006}, we
chose the values of $\alpha$ and $\beta$ in the range $2-5$ and $1-2$,
respectively, while two values of $T_{\rm peak}$ were explored ($10^{6.7}$\,K
and $10^7$\,K). Four sets of abundances were adopted: (i) the abundance set used
by \citet{GuedelXEST2006} to fit the spectra of known TMC members (hereafter
``TMC set''): with respect to the solar photospheric values by 
\citet{AndersGrevesse1989}, the abundances of this set are: C=0.45, N=0.788,
O=0.426, Ne=0.832, Mg=0.263, Al=0.5, Si=0.309, S=0.417, Ar=0.55, Ca=0.195,
Fe=0.195, Ni=0.195; (ii) all abundances equal to 0.2 times the solar
photospheric values of \citet{AndersGrevesse1989}; (iii) solar photospheric
values for all abundances, except for iron ($= 0.2$ solar); (iv) the ``TMC
set'', except for iron ($= 0.5$ solar). We finally considered two absorption
levels: ``low'' absorption with $N_{\rm H}=10^{21}$\,cm$^{-2}$, and ``high''
absorption with $N_{\rm H}=10^{22}$\,cm$^{-2}$. The fourteen input models are
listed in Table \ref{tab:modelli_input_simul}.
\begin{center}
\begin{table}[t]
\caption{Input models for simulations. ``TMC'' is the abundance set used by
\citet{GuedelXEST2006} (see text). ``Low'' and ``high'' absorption mean
$N_{\rm H}=10^{21}$\,cm$^{-2}$ and $N_{\rm H}=10^{22}$\,cm$^{-2}$,
respectively.}
\vspace{0.1cm}
\begin{center}
\begin{tabular}{lccccc} \hline\hline
Model &  $\log T_{\rm peak}$ & $\alpha$ &  $\beta$  & Abundances &  Absorption  \\ \hline
   1  &         7        &   3    &      2    &     TMC    &      low      \\
   2  &         7        &   3    &      2    &     TMC    &      high     \\
   3  &         7        &   5    &      2    &     TMC    &      low      \\
   4  &         7        &   5    &      2    &     TMC    &      high      \\
   5  &         7        &   2    &      1    &     TMC    &      low      \\
   6  &         7        &   2    &      1    &     TMC    &      high      \\
   7  &         7        &   3    &      2    & 0.2 solar  &      low      \\
   8  &         7        &   3    &      2    & 0.2 solar  &      high      \\
   9  &         7        &   3    &      2    & solar (Fe=0.2 solar) &  low  \\
  10  &         7        &   3    &      2    & solar (Fe=0.2 solar) &  high \\
  11  &         7        &   3    &      2    & TMC (Fe=0.5 solar)   &  low  \\
  12  &         7        &   3    &      2    & TMC (Fe=0.5 solar)   &  high \\
  13  &         6.7      &   2    &      1    &     TMC    &      low   \\
  14  &         6.7      &   2    &      1    &     TMC    &      high  \\ \hline
\end{tabular}
\end{center}
\label{tab:modelli_input_simul}
\end{table}
\end{center}

The two-temperature thermal models used to fit the simulated spectra are
based on the APEC emissivity code and have the following 13 free
parameters: the interstellar hydrogen column density ($N_{\rm H}$), the
temperature and emission measure of each plasma component, and the
abundances of O, Ne, Fe, Mg, Si, S, Ar, Ca. We fixed the abundances of C, N
and Ni to the respective values in the TMC set, because they are generally
not well-constrained from PN fitting; in fact, the C and N lines fall at
low energies where the PN resolution is quite poor (and moreover the PN
calibration is probably uncertain), while the Ni lines are blended with
the (usually) stronger iron and neon lines.

\begin{figure*}[t]
\begin{center}
\scalebox{0.64}{
\includegraphics{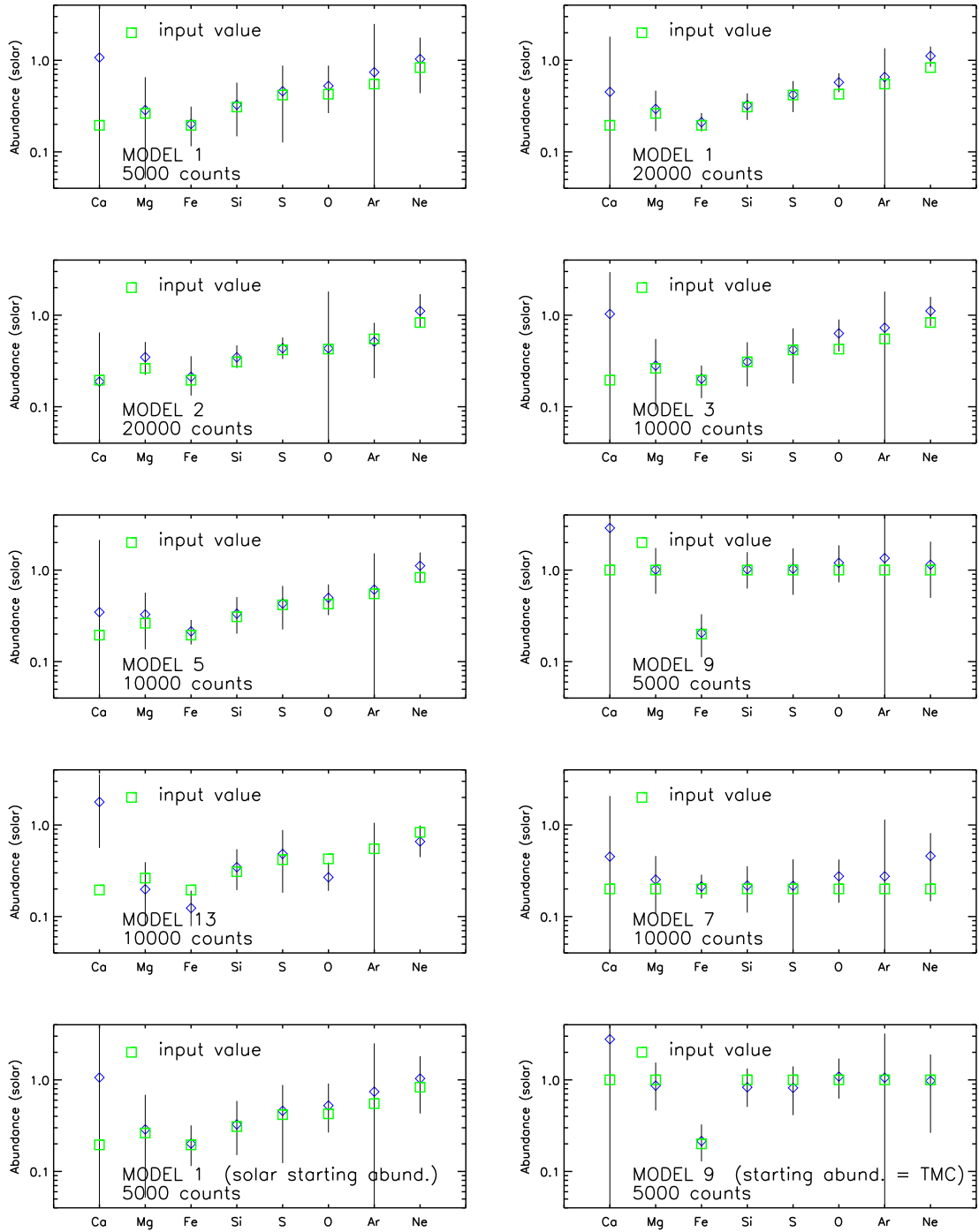}}
\caption{Medians (diamonds) and central 90\% of the distributions of the
abundances found from eight sample simulations, for different models and/or
statistics. The input values of abundances are shown as squares. The input EMDs
are equal in models 1, 2, 7 and 9, with $\log T_{\rm peak}=7$. The EMD of model
3 is steeper, while the EMD of model 5 is flatter. The EMD of model 13 peaks at
$\log T_{\rm peak}=6.7$. The input models have low absorption, with the
exception of model 2. The two bottom panels show the results obtained when the
starting abundances for the fitting are different from the ``input value'' of
the relevant model.}
\label{fig:esempio_ris_simul}
\end{center}
\end{figure*}

As representative examples of the results of all simulations, we
report in Fig. \ref{fig:esempio_ris_simul} those obtained for the abundances in
terms of medians of the distributions of the 1000 best-fit values, and error
bars corresponding to the central 90\% of the distributions, for ten
different cases illustrating the effects of different input models, statistics,
absorptions and choice of starting abundance values for the fittings.

The results indicated that, for all the cases analyzed here, the derived
abundances are generally in good agreement with the values of the input models.
The medians of the abundance distributions obtained from fitting simulated
spectra are, for all elements, compatible with the input value within 90\%
statistical errors, and within the 68\% errors in the great majority of the
cases. Some systematic effects must be noted in a few cases: Ne results
overestimated by a factor $\sim 2$ in models 7 and 11, while Fe and O are
underestimated by about the same factor in model 13.

As to the magnitude of the statistical uncertainties, errors at 90\%
confidence level within a factor $\sim$ 2.5 are obtained for most of the
elements using PN spectra with 5000 counts, and within a factor $\sim$ 2
with 10\,000 counts. Exceptions are oxygen, argon and calcium in some cases (see
below). Iron is the best-determined element, constrained within a factor
$\sim$ 1.6 when absorption is relatively low, already with spectra containing 
5000 counts. For the other elements, 90\% errors on the medians can be a
factor $\sim$ 1.4 up to 3 (spectra with 10\,000 counts), depending on the
model.

The abundance of oxygen is very poorly constrained (even for the highest
statistics) when the $N_{\rm H}$ is of the order of $10^{22}$\,cm$^{-2}$, as one
could expect because of the strong attenuation of the spectrum at low energies
($E \lesssim 1$\,keV; recall that the strongest lines of the He--like and
H--like ions of oxygen lie at $E \sim 0.6$\,keV). Regarding argon, its abundance
can be better constrained in the case of high absorption; this can be easily
understood by noting that, for fixed number of total counts, a larger fraction
of all the spectral counts goes in the Ar lines ($\sim 3$\,keV) when the
low-energy region of the spectrum is strongly suppressed by the interstellar
absorption. Instead, although the strongest Ca lines fall at $\sim 4.1$\,keV,
i.e. close to the Ar lines in the high-energy tail of the spectrum, the
abundance of this element is generally ill-constrained, unless the abundance is
of the order of the solar value, or higher, and the statistics are high. Since
the PN effective area is about the same at $\sim 3$\,keV  and $\sim 4.1$\,keV,
the difference with the case of Ar is likely due to the lower emissivity of the
calcium lines, compared to those of the Ar lines.

It is important to note also that the results of the simulations are
practically the same, in terms both of magnitude of the errors and agreement
between medians of the abundance distributions and input values, regardless of
the starting abundance values used to fit the spectra.

Finally, note that the simulations described in this section assumed no
background contamination, which may not be negligible at the highest
energies. For the bright stars discussed in this paper, we verified that only in
a couple of cases the background spectrum dominates over the source spectrum in
the energy range $E \gtrsim 5$\,keV, which was therefore excluded from the
fitting. To investigate the effect of the exclusion of the Fe {\sc xxv} 6.7\,keV
complex on the ability to constrain the iron abundance and the other parameters
of the model, we repeated the simulations for the models 1 and 2 by fitting
the simulated spectra in the energy range $0.3-5$\,keV. The results obtained in
these cases are very similar to those obtained employing a larger spectral
range.

\end{document}